\documentclass[useAMS,usenatbib]{mn2e}

\usepackage{amsmath}
\usepackage{epsfig}

\title[Simulating SMBH growth and feedback]{Cosmological simulations
  of the growth of supermassive black holes and feedback from active
  galactic nuclei: method and tests}

\author[C. M. Booth \& J. Schaye]{C. M. Booth$^{1}$\thanks{E-mail: booth@strw.leidenuniv.nl (CMB)} and Joop Schaye$^{1}$\\
$^{1}$Leiden Observatory, Leiden University, PO Box 9513, 2300 RA Leiden, the Netherlands}

\newcommand{\epsilonf}{\epsilon_{{\rm f}}}
\newcommand{\mbh}{m_{{\rm BH}}}
\newcommand{\nheat}{n_{{\rm heat}}}
\newcommand{\mg}{m_{{\rm g}}}

\begin{document}
\pagerange{\pageref{firstpage}--\pageref{lastpage}} \pubyear{2009}
\maketitle
\label{firstpage}
\begin{abstract}
We present a method that self-consistently tracks the growth of
supermassive black holes (BHs) and the feedback from active galactic
nuclei (AGN) in cosmological, hydrodynamical simulations. Our model is
a substantially modified version of the one introduced by
\cite{spri05} implemented in a significantly expanded version of the
     {\sc gadget III} code, which contains new prescriptions for star
     formation, supernova feedback, radiative cooling and
     chemodynamics. We simulate the growth of BHs from an initial seed
     state via Eddington-limited accretion of the surrounding gas, and
     via mergers with other BHs.  Because cosmological simulations at
     present lack both the resolution and the physics to model the
     multiphase interstellar medium, they tend to strongly
     underestimate the Bondi-Hoyle accretion rate. To allow low-mass
     BHs to grow, it is therefore necessary to increase the predicted
     Bondi-Hoyle rates in star-forming gas by large factors, either by
     explicitly multiplying the accretion rate by a numerical
     correction factor, or using an unresolved, subgrid model for the
     gas close to the BH. We explore the physical regimes where the
     use of such multiplicative factors is reasonable, and through
     this introduce a new prescription for gas accretion by
     BHs. Feedback from AGN is modeled by coupling a fraction of the
     rest-mass energy of the accreted gas thermally into the
     surrounding medium. We describe the implementation as well as the
     limitations of the model in detail and motivate all the changes
     relative to previous work. We demonstrate how general physical
     considerations can be used to choose many of the parameters of
     the model and demonstrate that the fiducial model reproduces
     observational constraints.

We employ a large suite of cosmological simulations, in which the
parameters of the BH model are varied away from their fiducial values,
to investigate the robustness of the predictions for the cosmic star
formation history and the redshift zero cosmic BH density, BH scaling
relations, and galaxy specific star formation rates. 
We find that the freedom introduced by the need to increase the
predicted accretion rates by hand, the standard procedure in the
literature, is the most significant source of uncertainty. Our
simulations demonstrate that supermassive BHs are able to regulate
their growth by releasing a fixed amount of energy for a given halo
mass, independent of the assumed efficiency of AGN feedback, which
sets the normalization of the BH scaling relations.  Regardless of
whether BH seeds are initially placed above or below the BH scaling
relations, they grow onto the same scaling relations. AGN feedback
efficiently suppresses star formation in high-mass galaxies.
\end{abstract}


%
%
\section{Introduction}
\label{sec:intro}

Over the past decades a growing body of observational and theoretical
evidence has suggested that supermassive black holes (SMBHs; ${\rm
  m}_{{\rm BH}}>10^6{\rm M}_{\odot}$) exist in the centers of all
galaxies with spheroids \citep[e.g.][]{korm95,ferr00} and that the
properties of these SMBHs are tightly correlated with the properties
of the spheroid in which they reside. For example, the mass of the
SMBH is found to be tightly correlated with the bulge stellar mass or
luminosity \citep{magg98,mclu02,hari04,laor01}, stellar velocity
dispersion \citep{gebh00,merr01,trem02}, and galaxy concentration, as
measured by the S\'{e}rsic index \citep{grah07}.  Some recent work has
demonstrated that these correlations may be understood in terms of a
black hole (BH) \lq fundamental plane\rq, relating BH mass, galaxy
effective radius, stellar velocity dispersion and stellar mass.  Here,
the mass of the SMBH essentially tracks the binding energy of the
stellar bulge \citep{marc03,feol05,alle07,hopk07}, although other
authors argue that the appearance of a fundamental plane is actually
due to biasing caused by the presence of galaxies with bars
\citep{grah08}.

The exact mechanisms leading to the tight observed coupling between
galaxy spheroidal components and central active galactic nuclei (AGN)
are not yet fully understood, although it has long been recognized
that the formation mechanisms of SMBHs \citep[e.g.][]{silk98} and
stars \citep[e.g.][]{deke86} are most likely self-regulating. These
results suggest that the same processes that shape galaxy spheroids
also act on the central BHs. Correlations between AGN activity and
other processes provide other clues about the mechanisms that lead to
the buildup of the SMBH population.  There is evidence that there
exists a link between galactic star formation and accretion onto a
central AGN: in a global sense, the evolution of the cosmic star
formation rate \citep[e.g.][]{mada96} and the luminosity density of
quasars are tightly correlated \citep{boyl98}.  Additionally, on the
scale of individual objects it has been found that the most powerful
narrow line AGN are preferentially found in galaxies that appear to
have undergone a recent starburst phase \citep{kauf03}.

The massive BHs present in the centres of galaxies are likely to have
started their lives as \lq seed\rq\, BHs. The typical masses of seed
BHs remains somewhat uncertain, and depends upon the mechanism by which
they form.  Plausible mechanisms include the collapse of population
III stars, giving rise to BHs with masses in the range $10^2{\rm
  M}_{\odot}<\mbh<10^3{\rm M}_{\odot}$
\citep[e.g.][]{mada01,isla03,schn02}, and direct collapse of matter in
high redshift, low angular momentum haloes, which may give rise to
seed BHs with masses $\sim10^5\,{\rm M}_\odot$
\citep[e.g.][]{loeb94,brom03,bege06,dijk08,volo09}. These seed mass
BHs can then grow either by mergers with other BHs
\citep[e.g.][]{isla03}, or through accretion of gas and/or stars.

Accretion of matter onto central BHs, accompanied by the release of a
fraction of the rest mass energy of the fuel, has long been recognized
as one of the most likely mechanisms to power AGN \citep{salp64}, and
a coupling between the accretion history of an AGN and the gas
dynamics of the bulge provides a plausible mechanism by which AGN and
bulge properties could become strongly correlated coupled
\citep[e.g.][]{silk98}.  For example, it has been suggested that the
central BHs grow until they release sufficient energy to unbind the
gas that feeds them from the host galaxy \citep{fabi99}.  Bursts of
AGN activity then expel gas from galaxies and remain quiescent until
stellar mass loss replenishes the galaxy's gas reservoir
\citep{ciot01}

A theoretical link between galaxy mergers and both galaxy-scale
starburst events and active AGN phases has been well established and
modelled.  Galaxy mergers have long been recognized as a mechanism by
which gas can potentially be channeled to the centre of a galaxy
\citep{toom72}, and N-body simulations of galaxy mergers confirmed and
extended this picture by showing that the asymmetrical gravitational
potential present during mergers is capable of funneling gas
efficiently to the center of a galaxy \citep{miho94}, where it may be
accreted by a SMBH.  Hydrodynamical simulations of galaxy mergers
\citep{barn91,barn96,miho96,kapf05}, and numerical models of AGN
growth \citep{kapf05,spri05a} predict that these merger events are
indeed responsible for the rapid growth of AGN. Recent numerical
studies \citep[e.g.][]{mici07} have indicated that both BH mergers and
gas accretion are important processes in forming the population of BHs
that we observe in the local universe.

Thus, it seems that we can paint a coherent picture in which emission
by AGN, galaxy mergers and the growth of supermassive BHs are closely
intertwined. As such the study of any one of these processes requires
an understanding of all of them.  For this reason detailed studies of
the co-evolution of the AGN and galaxy populations in a cosmological
context often resort to numerical techniques.  Early theoretical
studies of the galaxy-AGN connection relied upon dark matter halo
merger rates, without any separate treatment of galaxy formation
processes \citep{efst88,haeh93}.  Later work expanded upon this
groundwork by incorporating AGN feedback into semi-analytic modelling
of galaxy formation
\citep[e.g.][]{kauf00,catt01,bens03,gran04,crot06,bowe06,lago08,baug06}. Semi-analytic
models indicate that feedback from AGN is necessary in order to build
up a red-sequence of galaxies.  Although a combination of
photo-heating by reionization and supernova feedback can suppress star
formation in low mass haloes -- bringing the galaxy luminosity
function in line with observations at the low mass end -- models
without AGN feedback face the problem that the reheated gas in massive
haloes would eventually cool, giving rise to an excessive number of
bright galaxies \citep[e.g.][]{bowe06} as compared to the local
universe.

A more computationally challenging approach is to simulate galaxies
hydrodynamically, with additional sub-grid modelling of the growth and
energy feedback from AGN.  Numerical hydrodynamic simulations of
galaxy mergers containing AGN
\citep[e.g.][]{spri05,dima05,hopk06,robe06} have shown that the
presence of a central AGN can significantly alter the structure of
merger remnants, particularly by expelling a hot halo of diffuse,
low-angular momentum gas from the center of the remnant.  More recent
numerical studies have revealed that dissipation and dry mergers are
likely to play a fundamental role in shaping the co-evolution of BHs
and galaxies \citep{hopk08}.  Hydrodynamic simulations of full
cosmological volumes \citep{dima08,sija07,crof08,okam08} have probed
the effect of AGN on a cosmologically representative set of galaxies
and showed that the inclusion of AGN physics into galaxy formation
simulations allows us to match many of the observed properties of
galaxies in the local universe.  The modelling of an AGN population in
this manner is both computationally very expensive and subject to very
many, as yet poorly understood, numerical effects.  As such studies of
this type must take care to test the robustness of the models to all
physical and numerical parameters.

The focus of the current work is to present and test a new model for
the co-evolution of BHs and galaxies.  We note that nearly all BH
models published thus far in the literature employ the star formation
and supernova feedback models of \citet{spri03} (hereafter SH03), and
the model for BH growth and AGN feedback of
\citet{spri05}\footnote{Although see \cite{okam08} for a different
  approach.} (hereafter S05).  Throughout this paper we highlight
similarities and differences between our approach and that used
previously in the literature.  The primary difference between our
models and previous work is that we employ a different parametrization
of the process of gas accretion onto BHs as well a different
implementation of AGN feedback.  We show that changes to the BH
accretion model can lead to profound differences in galaxy properties,
global star formation rates and BH demographics. We examine both the
global properties of the simulation, such as the integrated star
formation rate and cosmic BH density, and consider the properties of
individual galaxies, including specific star formation rates, and the
BH fundamental plane.  We quantify how uncertainties in our numerical
model and all of our parameter choices affect the reliability of our
results.  We find that changes in the numerical model that generates
seed mass BHs, and in the model that distributes feedback energy into
the ISM do not strongly affect our results.  However, the accretion
model is found to be of crucial importance in understanding our
results.  Throughout we assume a flat $\Lambda$CDM cosmology with the
cosmological parameters:
$\{\Omega_m,\Omega_b,\Omega_\Lambda,\sigma_8,n_s,h\}=\{0.238,0.0418,0.762,0.74,0.951,0.73\}$,
as determined from the WMAP 3-year data \citep{sper07} and
consistent\footnote{Our value of $\sigma_8$ is 1.6$\sigma$ lower than
  allowed by the WMAP 5-year data.} with the WMAP 5-year data
\citep{koma08}.  Where necessary, observational results have been
scaled to our chosen cosmology, and the stellar initial mass function
(IMF) assumed in observational analyses has been scaled to the
Chabrier IMF used in our simulations.

The paper is structured as follows: In Sec.~\ref{sec:method} we
introduce our simulation set, and describe briefly the sub-grid
physics modules that are not directly related to BHs.  In
Sec.~\ref{sec:met-bh} we describe in detail our model for BH
formation, growth and feedback and we motivate our choices for
numerical parameters in Sec.~\ref{sec:pars}.  In Sec.
\ref{sec:results} we present simulation results, including a
comparison with redshift zero observational data and an investigation
into the severity of uncertainties introduced by different parameter
choices.  Finally, in Sec.~\ref{sec:discussion} we discuss and
summarize our findings.  In a companion work we investigate in detail
the interplay between feedback from AGN and other feedback processes,
including winds driven by Type 2 supernovae and mass loss from the
stellar population.

%
%
\section{Numerical simulations}
\label{sec:method}
In this section we introduce the numerical techniques used in our
simulations and provide a brief overview of the sub-grid physics
modules that are not directly related to BH growth or AGN feedback.
  
We have carried out a suite of cosmological simulations using
Smoothed-Particle Hydrodynamics (SPH) \citep{lucy77,ging77,mona92},
employing a significantly extended version of the parallel PMTree-SPH
code {\sc gadget III} \citep{spri05a,spri01}, a Lagrangian code used
to calculate gravitational and hydrodynamic forces on a
particle-by-particle basis.  The initial particle positions and
velocities are set at $z=127$ using the Zel'dovich approximation to
linearly evolve positions from an initially glass-like state. The
production simulations used in this study are run in boxes of size 50
comoving Mpc/$h$, and contain $256^3$ particles of both gas and dark
matter.  Comoving gravitational softenings are set to $1/25$ of the
mean comoving inter-particle separation down to $z=2.91$, below which
we switch to a fixed proper scale of 2~kpc$/h$.  The production
simulations have gas particle masses of $8.64\times10^{7}\,{\rm
  M}_\odot/h$. The boxes are evolved all the way to redshift zero.

In addition to hydrodynamic forces we treat star formation, supernova
feedback, radiative cooling, chemodynamics, black hole accretion, and
AGN energy feedback in these simulations.

Star formation is tracked in the simulations following the
prescription of \citet{scha08}.  Gas with densities exceeding a
critical density for the onset of the thermo-gravitational instability
(hydrogen number densities $n_{{\rm H}}=10^{-2}-10^{-1}$ cm$^{-3}$) is
expected to be multiphase and to form stars \citep{scha04}. Because we
lack both the physics and the resolution to model the cold
interstellar gas phase, we impose an effective equation of state (EOS)
with pressure $P\propto \rho^{\gamma_{{\rm eff}}}$ for densities
$n_{{\rm H}} >n_{{\rm H}}^*$ where $n_{{\rm H}}^*=0.1$~cm$^{-3}$,
normalised to $P/k=10^3$~cm$^{-3}$K at the threshold.  We use
$\gamma_{{\rm eff}}=4/3$ for which both the Jeans mass and the ratio
of the Jeans length to the SPH kernel are independent of the density,
thus preventing spurious fragmentation due to a lack of numerical
resolution \citep{scha08}.  As described in \cite{scha08}, gas on the
effective EOS is allowed to form stars at a pressure-dependent rate
that reproduces the observed Kennicutt-Schmidt law \citep{kenn98} by
construction, renormalised by a factor\footnote{This normalization
  factor is calculated from the asymptotic ratio of the numbers of
  ionizing photons predicted from models of stellar populations with a
  constant SFR \citep{bruz03}.} 1/1.65 to account for the fact that it
assumes a Salpeter IMF whereas we use a Chabrier IMF.

Energy injection due to supernovae is included through kinetic
feedback.  We employ the prescription of \cite{dall08}, which is a
variation of the SH03 recipe for kinetic feedback.  In this
prescription core-collapse supernovae locally inject kinetic energy
and kick gas particles into winds.  The feedback is specified by two
parameters: Firstly, the initial mass-loading $\eta=\dot{m}_{\rm
  w}/\dot{m}_*$, which describes the initial amount of gas put into
the wind, $\dot{m}_{\rm w}$, as a function of the local SFR,
$\dot{m}_*$, and secondly the wind velocity, $v_{{\rm w}}$.  We use
$\eta=2$ and $v_{\rm w}=600$~km/s, which corresponds to 40\% of the
total amount of supernova energy.  In contrast with the models of
SH03, the kinetic energy is injected \emph{locally} to every star
formation event and wind particles are \emph{not} temporarily
decoupled from the hydrodynamics when they are put into the wind.

As described in \cite{wier09}, we follow the timed release of 11
different elements from massive 
stars (Type II supernovae and stellar winds) and intermediate mass
stars (Type Ia supernovae and asymptotic giant branch
stars), assuming a Chabrier initial mass function spanning the range
0.1 to 100~${\rm M}_\odot$.  
Radiative cooling was implemented following
\cite{wier08}\footnote{We used their equation (3) rather than (4) and
  {\sc cloudy} version 05.07 rather than 07.02.}. In brief, net
radiative cooling rates 
are computed element-by-element in the presence of the cosmic
microwave background and a \cite{haar01} model for the UV/X-ray
background radiation 
from quasars and galaxies. The contributions of the eleven elements
hydrogen, helium, carbon, nitrogen, oxygen, neon, magnesium,
silicon, sulphur, calcium, and iron are interpolated as a function of
density, temperature, and redshift from tables that have been precomputed
using the publicly available photo-ionization package
{\sc CLOUDY}, last described by \cite{ferl98}, assuming the gas
to be optically thin and in (photo-)ionization equilibrium.

%
%
\section{The black hole model}
\label{sec:met-bh}

We now provide a detailed description of our models for BH formation
and accretion (Sec.~\ref{sec:met-accr}), BH mergers
(Sec.~\ref{sec:met-merger}) and energy feedback from AGN
(Sec.~\ref{sec:met-feed}).  Throughout this section we highlight and
justify the aspects of our model that differ from previous works.  We
also introduce all the relevant parameters.  In Sec.~\ref{sec:pars} we
motivate our choices for these parameters.

\subsection{Black hole formation and accretion}
\label{sec:met-accr}

Plausible BH seed formation mechanisms lead to the creation of BHs
with masses in the range $10-10^5\,{\rm M}_\odot$, whereas SMBHs in
the local Universe have masses of up to $10^9\,{\rm M}_\odot$ (see Sec
\ref{sec:intro} for a discussion). To understand the origin of the
redshift zero BH population we therefore need to model how BHs can
grow to the sizes observed in present-day galaxies.  Over the past
decades a picture has emerged in which SMBHs are embedded in dense
stellar systems in the centres of galaxies and increase their masses
primarily by the accretion of gas \citep[e.g.][]{bege78}.  BHs may
also grow by mergers with other BHs, or by the disruption and capture
of stars \citep[e.g.][]{lynd69}.  The capture of stars has been put
forward as an explanation for ultra-luminous X-ray sources \citep[see
  e.g.][for a review]{fabb06}. However, we neglect this process in the
current work, and instead focus on how BHs can accrete gas from their
surroundings.

The model presented in this section is a substantially modified
version of the model 
introduced by S05 and employed in almost all of the large-scale
numerical simulations of AGN growth thus far available in the
literature (see Table \ref{tab:allparlist} for an overview).

Because cosmological simulations have neither the resolution nor the
necessary physics to simulate the formation of the seed BHs that
eventually grow into SMBHs, it is assumed that low-mass seed BHs are
produced sufficiently regularly that every halo above a certain
threshold mass contains one such object at its center.  Here, our
model follows that of \cite{dima08} in that we regularly run a
friends-of-friends group finder with linking length equal to 0.2 times
the initial mean inter-particle spacing \citep{davi85} on all of the
dark matter particles during the simulation. We do so at times spaced
evenly in log expansion factor, $a$, such that $\Delta a=0.02a$, which
corresponds to a proper time of $\sim$250~Myr ($\sim 70\,{\rm Myr}$)
at redshift zero (three) for our cosmology.  When a halo grows above
some threshold mass, $m_{{\rm halo,min}}$, and does not already
contain a BH, then its most gravitationally bound baryonic particle is
converted into a collisionless BH particle.  The initial mass of these
BHs is usually chosen to be well below the resolution limit of our
cosmological simulations (see Sec.~\ref{sec:pars}), and as such we
need to employ sub-grid models to follow the BH.  Although we convert
the entire particle into a \lq black hole particle\rq, the mass of the
seed BH ($m_{{\rm seed}}$) associated with this particle is usually
initially significantly less than the particle mass ($m_{{\rm seed}}
\ll m_{{\rm g}}$; where $\mg$ is the simulation gas particle mass).
We therefore store the mass of the subgrid BH separately.  For the
gravitational interactions, other than BH accretion, the full mass of
the particle ($\mg$) is used, but for calculating the BH-specific
processes we use the sub-grid BH mass ($m_{{\rm BH}}$).  We now
discuss in more detail the manner in which we track the growth of the
BH.
 
BH particles are collisionless sink particles that contain a sub-grid
BH, initially of mass $m_{{\rm seed}}$, chosen to be well below the
observed mass of BHs in haloes of this size.  From their initial seed
mass, BHs may grow via one of two processes: mergers with other BHs
and accretion of surrounding ambient gas.  We now treat each of these
processes in turn.  In our models BHs accrete from the surrounding
ambient gas phase at a rate proportional to that given by the
Bondi-Hoyle-Lyttleton \citep{bond44,hoyl39} formula
\begin{equation}
 \dot{m}_{{\rm accr}}=\alpha\frac{4\pi G^2 \mbh^2 \rho}{(c_{s}^2+v^2)^{3/2}}\,,
 \label{eq:bhl}
\end{equation}
where $\mbh$ is the mass of the BH, $c_{s}$ and $\rho$ are the sound
speed and gas density of the local medium, $v$ is the velocity of the
BH relative to the ambient medium, and $\alpha$ is a dimensionless
efficiency parameter. The factor $\alpha$ did not appear in the
original analyses of \citet{bond44} and \citet{hoyl39}, but was
introduced by S05 as a numerical correction factor, to compensate for the
limitations of the numerical simulations.  The assumption that BHs
grow via Bondi-Hoyle accretion is reasonable even if they are in
reality fed by accretion discs that are far smaller than the
resolution limit of our simulations as long as the latter grow by
Bondi-Hoyle accretion. However, we will see that very large
factors of $\alpha$ are required for low-mass BHs to grow, in which
case one cannot claim to be simulating Bondi-Hoyle accretion.

 The amount of accreted mass is related to the rate of growth of the
 BH by\footnote{Note that S05 neglected the $(1-\epsilon_{\rm r})$
   term and used $\dot{m}_{{\rm BH}}=\dot{m}_{{\rm accr}}$.}
 $\dot{m}_{{\rm BH}}=\dot{m}_{{\rm accr}}(1-\epsilon_{\rm r})$, where
 $\epsilon_{\rm r}$ is the radiative efficiency of a BH, which we
 always assume to be 10\%, the mean value for the radiatively
 efficient \cite{shak73} accretion onto a Schwarzschild BH.

In order to resolve Bondi-Hoyle accretion onto a BH we need to resolve
the Bondi-Hoyle radius ($r_{\rm b}$), defined as
\citep[e.g.][]{edga04}:
\begin{equation}
r_{{\rm b}}=\frac{G\mbh}{c_s^2}\approx 0.042\Big(\frac{M_{{\rm BH}}}{10^6\,{\rm M_\odot}}\Big)\Big(\frac{c_s}{10\,{\rm km/s}}\Big)^{-2}\,{\rm kpc}.
\end{equation}
Comparing this to the Jeans length,
\begin{equation}
L_{\rm J}\sim \sqrt{\frac{c_s^2}{G\rho}}\sim\frac{GM_{{\rm J}}}{c_{{\rm s}}^2}\,,
\end{equation}
where $M_{{\rm J}}$ is the Jeans mass, we see that $r_{{\rm
    b}}\sim L_{{\rm J}}$ if $\mbh\sim M_{{\rm J}}$ and $r_{{\rm b}}\gg
L_{{\rm J}}$ if $\mbh\gg M_{{\rm J}}$.  Hence, any simulation
that resolves the Jeans scales will also resolve accretion onto black
holes of mass $\mbh > \mg$. We can then parametrize accretion onto a BH in two different
ways:

{\bf 1. Density-Independent Accretion Efficiency:} Most AGN models in
the literature use a constant value of $\alpha = 10^2$
\citep[e.g.][see also
  Table~\ref{tab:allparlist}]{spri05,dima05,sija07,dima08,bhat08,colb08,crof08,joha09}. Although
most authors do not motivate or even mention \footnote{
  \cite{spri05,dima05,sija07,dima08,bhat08,colb08} all do not discuss
  or mention the value of $\alpha$ that they used, but we have been
  informed by V.~Springel that they assumed
  $\alpha=100$. \citet{khal08} state explicitly that in their models
  $\alpha=300$ and justify this by noting that when the density of the
  ISM is smoothed on the scale of the computational resolution, the
  recovered densities are much lower than would be expected on the
  scale of the Bondi radius.  Using similar reasoning, \cite{joha09}
  reach a similar conclusion and set $\alpha=100$.} their choice of
$\alpha$, we note that values much greater than unity can be justified
in one of two ways: firstly by noting that the Bondi-Hoyle accretion
rate depends strongly upon the local ISM sound speed.  Galaxy
formation simulations currently have neither the resolution nor the
physics to self-consistently track the properties of the cold-phase of
the ISM, and as such the temperature of the gas accreted by the AGN
may be overestimated by orders of magnitude.  Hence, we can justify
very large values of $\alpha$ in star forming gas. Secondly, in
low-resolution simulations we do not resolve the Jeans scale, even in
single-phase gas, so the density of the gas at the Bondi radius is
underestimated, allowing us to again justify large values of $\alpha$.
We call models that use a fixed value of $\alpha$ \lq
constant-$\alpha$\rq\, models.  Constant-$\alpha$ models are
parametrized by a constant multiplicative factor in the Bondi-Hoyle
accretion rate, $\alpha_0$.  An alternative way of increasing
accretion rates is to employ a subgrid model for the unresolved ISM
properties, and use this to artificially boost the ISM densities local
to the BHs.  We discuss this further in the following sections.

We will show in Sec.~\ref{sec:bhg} that the use of a constant-$\alpha$
model has a profound effect on the ability of BHs to grow, and that
changes in this very poorly constrained parameter can lead to large
changes in the global properties of the simulation such as the global
density in BHs. We emphasize that values of $\alpha \gg 1$ imply the
assumption that the simulation predictions for the gas density and
temperature are sufficiently wrong that the Bondi-Hoyle accretion rate
is underestimated by two orders of magnitude. Although this assumption
can be justified for high-density gas, it does mean that the value of
$\alpha$ is in fact more important than the predicted densities and
temperatures. One could therefore argue that models of this kind do
not really simulate Bondi-Hoyle accretion.

Cosmological simulations can, however, already model Bondi-Hoyle
accretion of low-density gas and it hence makes sense to use an
accretion model for which the ``fudge factor'' $\alpha$ becomes unity
in the regime where the simulations are reliable. We therefore 
introduce a new class of black hole accretion models in which the
the value of $\alpha$ depends on the local gas density, while
keeping the number of free parameters fixed.

{\bf 2. Density-Dependent Accretion Efficiency:} The assumptions used
to justify large values of $\alpha$ in simulations similar to ours
break down when two conditions are satisfied: firstly the local gas
density must be lower than required for the formation of a cold
(i.e.\ $T\ll 10^4$~K) phase, and secondly the simulation must resolve
the Jeans scale of the single-phase gas.  Our highest resolution
simulations (as well as many published AGN simulations) do resolve the
Jeans scale at the star formation threshold and so in contrast to most
published AGN models we choose to parametrize the accretion
efficiency parameter as a function of density

\begin{equation}
\label{eq:beta}
 \alpha=\left\{ \begin{array}{cc}
    1 & {\rm if}\,n_{{\rm H}}<n_{{\rm H}}^* \\
    \Big(\frac{n_{{\rm H}}}{n_{{\rm H}}^*}\Big)^\beta &
    \textrm{otherwise.}\end{array}\right.
\end{equation}
Here, the accretion efficiency ($\alpha$) becomes unity for densities
lower than the critical value required for the formation of a cold
interstellar gas phase ($n_{{\rm H}}^*=0.1\,{\rm cm}^{-3}$; see
Sec.~\ref{sec:method}).  As discussed above, provided the simulations
resolve the Jeans scale, the Bondi radius will be resolved for BHs
with $\mbh \ge \mg$, which means that values of $\alpha \gg 1$ are
unphysical for such BHs.  We then choose to parametrize our lack of
knowledge about both the physical properties of the multiphase ISM and
the rate at which it accretes onto the central AGN using a power-law
of the gas density, with slope $\beta$.  This constant-$\beta$ model,
which has the same number of free parameters (one) as the
constant-$\alpha$ models used in previous work, allows us to correctly
describe accretion in the physical regime where it is resolved by our
simulations and to introduce a reasonable scaling when it is not.  We
will show in Sec.~\ref{sec:res-pareffects} that the change from a
constant-$\alpha$ to a constant-$\beta$ model can have a profound
effect on the growth of BHs, particularly for low mass galaxies.  We
call models of this type \lq constant-$\beta$ models\rq.
 
A second approach to boosting accretion rates, which operates in a
similar manner to the constant-$\beta$ models, is to make use of a
subgrid model for the unresolved subgrid physics not encapsulated by
the simulations.  For example, \citet{pelu07} use the star formation
and supernova feedback models of SH03 to estimate the amount of time
that a BH spends in dense, molecular clouds, and \citet{okam08} use a
subgrid model in which drag due to stellar radiation on a clumpy ISM
can give rise to large accretion rates onto a central BH.  We note
that differing implementations of the subgrid model can lead to large
differences in the properties of the ISM and for the purposes of this
work we emphasize that the functional form (Eq.~\ref{eq:beta}), as
well as the value for $\beta$, are ad-hoc. Any function for which
$\alpha \rightarrow 1$ at gas densities for which the simulations are
reliable and for which $\alpha \gg 1$ at higher densities would do. We
chose to use a simple power-law dependence because it satisfies these
constraints, is continuous and uses only one free parameter. We will
investigate the effect of changing $\beta$ in the following
sections. In the limit that $\beta \rightarrow 0$ the behaviour of the
constant-$\beta$ model will tend towards that of a constant-$\alpha$
model with $\alpha_0=1$, and in the limit that $\beta \rightarrow
\infty$ the model tends towards behaviour where the accretion is pure
Bondi-Hoyle in non star-forming gas, and always Eddington limited in
gas with densities above the star formation threshold.  We caution
that this prescription is not suitable for simulations that resolve
the relevant physics at densities exceeding $n_{\rm H}^\ast$.

Because we have changed the density-dependence of the
accretion rate, we cannot claim to be simulating Bondi-Hoyle
accretion. Values of $\alpha\gg 1$ are, however, motivated by the
Bondi-Hoyle formula. Moreover, for $n_{\rm H} > n_{\rm H}^\ast$ the density 
should be interpreted as the mass-weighted mean density of the
unresolved, multiphase medium, smoothed on the scale of the spatial
resolution of the simulation, whereas the density appearing in the
Bondi-Hoyle formula applies to a single gas phase. Since the
accretion rate-weighted mean density (which we can only compute if we
know the mass distribution of the multiphase gas as a function of
density and temperature) is unlikely to be proportional to this
effective density, there is no reason to keep the Bondi-Hoyle
scaling. For this reason, and because $\alpha\gg 1$ implies the
assumption that the predicted densities and temperatures are greatly
in error, we argue that constant-$\alpha$
prescriptions with $\alpha \gg 1$ can no more claim to be modeling
Bondi-Hoyle accretion than constant-$\beta$ prescriptions. We prefer
the latter since it allows us to get the right answer in the regime
where the simulations are reliable, that is, at sufficiently low
densities. Finally, we note that both accretion models use the
Bondi-Hoyle scaling of the accretion rate with the mass of the BH, 
$\dot{m}_{\rm accr} \propto m_{\rm BH}^2$.

In common with the models of S05 we limit the accretion rate to the
Eddington rate:  
\begin{equation}
 \dot{m}_{{\rm Edd}}=\frac{4\pi G \mbh m_{{\rm p}}}{\epsilon_{{\rm r}} \sigma_{{\rm T}} c}\,,
\end{equation}
where $m_{{\rm p}}$ is the proton mass and $\sigma_{\rm T}$ is the
Thomson cross section for scattering of free electrons.  Because
$\dot{m}_{{\rm Edd}}\propto \mbh$ whereas $\dot{m}_{{\rm
    accr}} \propto \mbh^2$, Eddington limited accretion tends
to be more important for more massive BHs.

Following S05, we allow BH particles to stochastically swallow
neighbouring baryonic particles with a probability
 \begin{displaymath}
 p_i=\left\{ \begin{array}{ll}
    (\mbh-m_{{\rm part}})\rho^{-1}W(r_{{\rm BH}}-r_i,h_{{\rm BH}}) & \textrm{if } \mbh>m_{{\rm part}} \\
    0                            & \textrm{otherwise}\end{array}\right.
 \end{displaymath}
where $\rho$ is the local gas density, $\mbh$ is the mass of the
sub-grid black hole, $m_{{\rm part}}$ is the mass of the particle
  containing the sub-grid BH and $W(r_{{\rm BH}}-r_i,h_{{\rm BH}})$ is
  the SPH kernel, evaluated between the positions of the BH and gas
  particle $i$.  The BH smoothing length, $h_{{\rm BH}}$, is chosen
  such that within a distance $h_{{\rm BH}}$ from the BH there are
  $N_{{\rm ngb}}=48$ neighbours, the same number of neighbours as we
  used in our SPH calculations.  This process ensures that the mass of
  the BH particle always closely tracks $m_{{\rm BH}}$

When the mass of the BH particle is smaller than or of the same order
of magnitude as the simulation mass resolution, the black hole does
not dominate the local dynamics and may wander from the centre of mass
of its parent halo due to numerical effects.  Conservation of momentum
from accreted ISM gas can lead to similar effects.  In order to avoid
this we employ the same scheme as in the models of S05.  At every
timestep the gravitational potential energy is calculated at the
position of each of the BH's neighbouring gas particles and the BH
particle is repositioned on top of the particle with the minimum
potential energy.  In order to prevent the BH from being \lq
dragged\rq\, by a minimum-potential particle with a large relative
velocity, we only perform this process if the relative velocity
between the BH and its most-bound gas particle neighbour is less than
0.25~$c_s$, where $c_s$ is the local sound speed.  This process
ensures that the location of the BH particle always tracks the centre
of mass of its parent halo very closely.  This procedure is halted
after the mass of the SMBH becomes greater than ten times the initial
gas particle mass in the simulation because by this point the BH
dominates the dynamics in the centre of the halo.

\subsection{Black hole mergers}
\label{sec:met-merger} 
Galaxy mergers are thought to be one of the major processes driving
the evolution of galaxies.  When galaxies merge it is expected that
their central BHs will eventually also merge. Indeed, the build-up of
BHs through mergers may play an important part in the growth of SMBHs.
Similarly to S05 we have implemented BH merging as follows. When any
two BHs pass within a distance $h_{{\rm BH}}$ of each other with a
relative velocity smaller than the circular velocity at a distance
$h_{{\rm BH}}$ ($v_{{\rm rel}}<\sqrt{Gm_{{\rm BH}}/h_{{\rm BH}}}$,
where $h_{{\rm BH}}$ and $\mbh$ are the smoothing length and mass of
the most massive BH in the pair respectively) then they are allowed to
merge.  This velocity criterion is necessary in order to prevent BHs
from merging during a fly-through encounter of two galaxies, as this
could lead to BHs being quickly removed from their host galaxies due
to momentum conservation. This velocity scale is somewhat different
from that employed by S05, who used the local sound speed, $c_s$, as
the relevant velocity scale, arguing that the sound speed represents a
simple measure of the characteristic velocity scale of the galaxies,
and hence gives a simple measure of the velocity scale at which BHs
will be able to merge.  However, because AGN input large amounts of
energy into their surroundings, it is not necessarily true that the
sound speed local to the AGN reflects the depth of the potential well.

The BH merging rate estimated from our simulations likely represents
an upper limit to the true merger rate as our simulations do not have
the resolution required to resolve the formation of the tight BH
binaries that are a prerequisite for their eventual coalescence
\citep{call08}.  Since it is not yet fully understood how long it
takes to harden a BH binary \citep{maki04}, we assume that the merging
process is instantaneous.

\subsection{Energy feedback from black holes}
\label{sec:met-feed}

The precise mechanism by which energy emitted from a BH is coupled to
the surrounding medium is as yet unknown, but plausible mechanisms
include radiation pressure on free electrons (which gives rise to the
classical Eddington limit), Compton heating of the infalling gas
\citep[e.g.][]{ciot01,wang06b}, photoionization pressure
\citep{buff74,cowi78} and radiation pressure on dust grains
\citep[e.g.][]{murr05}.  Regardless of the precise coupling mechanism,
there is a catalogue of observational evidence indicating that energy
output from AGN can drive galactic outflows. For example, absorbers
seen in X-rays show evidence of outflow \citep{laor97} and broad
absorption line systems show evidence of outflows at very high
velocity \citep[e.g.][]{poun03}.  Although these observations indicate
that high velocity outflows are present around some AGN, they do not
tell us how much mass (and hence how much energy) is present in the
outflow.  Estimates of the mass outflow rate in the winds are highly
uncertain.  Some studies \citep[e.g.][]{chel08} imply that the actual
rate of mass outflow is only a small fraction of the bolometric
luminosity of the AGN sources, while other studies
\citep[e.g.][]{arav02,arav08} suggest large mass outflow rates in
quasar driven winds.

In our models BHs inject a fixed fraction of the rest mass energy of
the gas they accrete into the surrounding medium.  The feedback is
implemented thermally, that is: energy is deposited into the
surrounding gas by increasing its internal energy, as opposed to the
kinetic feedback used to inject supernova energy, which is deposited
by kicking the gas particles (see Sec.~\ref{sec:method}).  The
fraction of the accreted rest mass energy that is injected is assumed
to be independent of both the environment and the accretion rate.  We
thus do not differentiate between \lq quasar mode\rq\, and \lq radio
mode\rq\, feedback as in the models of \cite{sija07}. In a future work
we will consider how spatially distributed AGN heating mechanisms
affect the cosmological evolution of galaxies.  The amount of energy
returned by a BH to its surrounding medium in a timestep $\Delta t$ is
given by
\begin{equation}
\label{eq:epsilon}
 E_{{\rm feed}}=\epsilonf \epsilon_{\rm r} \dot{m}_{{\rm BH}} c^2 \Delta t\,,
\end{equation}
where $\epsilonf$ is the efficiency with which a BH couples the
radiated energy into its surroundings -- a free parameter in our
simulations -- and $c$ is the speed of light.  Only the product of
$\epsilon_{{\rm r}}$ and $\epsilonf$ is important in calculating the
amount of energy feedback in our model.

Because our sub-grid model for SF relies on an effective EOS and does
not include a (semi-)analytic sub-grid model for the multiphase ISM,
our energy distribution mechanism is different from that in S05.  In
contrast, because we prefer to minimize the use of semi-analytic
models within our hydrodynamical simulations, our models rely only on
an effective EOS and leave the distribution of the mass over
unresolved gas phases undefined.  We therefore need to make two
changes to the EOS model of \cite{scha08} that was used in our
simulations without AGN feedback.

Firstly, in the original models, once gas was identified as
star-forming it was forced to remain on the EOS, until its density
dropped below the critical density for star-formation, $n_{{\rm
    H}}^*$, it turned into a star particle, or it was kicked into the
wind.  It is therefore necessary that we change this by allowing
strongly heated gas to leave the EOS.  This is implemented numerically
by taking gas that is heated by more than 0.5~dex above the EOS in a
single time step off the EOS (i.e.\ it is no longer star-forming and
its pressure is no longer constrained to lie on the EOS).  Gas is
placed back onto the EOS if its temperature falls back below 0.5~dex
above the EOS temperature corresponding to its density.  By checking
SFRs, both globally and for individual objects, and by comparing gas
distributions on the $\rho-T$ plane we have verified that making this
change to our EOS model has a negligible effect on the results in a
simulation that does not include AGN feedback.  A second possible
change to the AGN model would have been to treat the effective EOS as
a lower limit to the gas temperature. We tested this and again found
the differences in our results to be negligible. We choose to use the first
procedure in order to facilitate direct comparisons between the simulations
containing AGN and those run earlier in the project.

Secondly, in order to ensure that the thermal feedback from BHs is not
immediately radiated away it is necessary to impose a \emph{minimum
  heating temperature}.  BHs store feedback energy until they have
accumulated enough energy to increase the temperature of $\nheat$ of
their neighbours by an amount of $\Delta T_{{\rm min}}$, i.e.
\begin{equation}
E_{{\rm crit}}=\frac{\nheat \mg k_{{\rm B}} \Delta T_{{\rm min}}}{(\gamma-1)\mu m_{{\rm H}}}\,,
\end{equation}
where $E_{{\rm crit}}$ is the critical energy for a heating event to
be triggered and $\mu$ is the mean molecular weight of the gas (we
assume $\mu=0.58$, appropriate for a fully ionized gas of primordial
composition).  The internal energy of the heated gas is
instantaneously increased by an amount $E_{{\rm crit}}$.  This
implementation of quasar mode feedback is similar to the radio mode
feedback introduced by \citet{sija07}. If $\Delta T_{{\rm min}}$ is
set too low then the cooling time of the AGN heated gas will remain
very short, and the energy will be efficiently radiated away.  If
$\nheat\Delta T_{{\rm min}}$ is set too high then the threshold energy
for a heating event to occur and hence the time period between AGN
heating events will become very large.  In particular, a time interval
larger than the Salpeter time would prevent the BH from regulating its
growth.  Finally, we note that the energy is deposited into the
ambient gas isotropically, equally distributed to a random fraction
$n_{{\rm heat}}/N_{{\rm ngb}}$ of the BH's neighbours.  If, on a given
timestep, a BH accretes more energy than necessary to heat $n_{{\rm
    heat}}$ particles to $\Delta T_{{\rm min}}$ then the process it
repeated until the BH has distributed all of its energy, so individual
gas particles may be heated by an amount $\Delta T_{{\rm min}}$
multiple times on a given timestep.

%
%
\section{Parameter choices}
\label{sec:pars}
Both the mechanism by which BHs grow and the efficiency of
their thermal feedback can be changed drastically by changing the
values of the parameters of the AGN model.  In this section we discuss
how parameter values are chosen to minimize unphysical numerical effects
whilst simultaneously requiring that the global properties of the
BH distribution satisfy various observational constraints.
For quick reference, Table \ref{tab:parlist} contains a full list of the
parameters that control the behaviour of the BH growth and AGN
feedback model, along with their fiducial values.

%
%
\begin{table*}\begin{center}
\caption{Full list of parameters of the AGN model, along with the
  values they take in the fiducial set of simulations, and short definitions.}
\begin{tabular}{l|l|l}
Parameter&Fiducial Value&Description\\
\hline
$m_{{\rm seed}}$&$0.001\,\mg$&Initial mass of the sub-grid BH\\
$m_{{\rm halo,min}}$&$100\,m_{{\rm DM}}$&Minimum halo mass into which BH seeds may be placed\\
$\epsilon_{\rm r}$&0.1&Radiative efficiency of the BH accretion discs\\
$\epsilon_{\rm f}$&0.15&Fraction of energy emitted by BHs that couples into the ambient gas\\
$\nheat$&1&Number of neighbouring particles heated per feedback event\\
$\Delta T_{{\rm min}}$&$10^8$~K&Amount by which BH feedback heats surrounding gas\\
\multicolumn{3}{|c|}{Depending on the accretion model, one of the following parameters is used:}\\
$\alpha_0$&100&Normalization of the Bondi-Hoyle accretion efficiency (Eq.~\ref{eq:bhl}) \emph{in constant-$\alpha$ models}\\
$\beta$&2&Slope of the Bondi-Hoyle accretion efficiency (Eq.~\ref{eq:beta}) \emph{in constant-$\beta$ models}\\
\end{tabular}
\label{tab:parlist}
\end{center}\end{table*}

Because it is difficult to discuss the effect of each parameter in
isolation, we will first discuss the general properties of the BH model
(e.g.\ growth mechanisms, feedback efficiency) and use each of these
general themes to motivate our fiducial choices for the parameters of
the AGN model.

\subsection{Black hole growth}
\label{sec:bhg}
We allow BHs to grow by two processes: mergers with other BHs and
accretion of gas.  The BH accretion time-scale $t_{{\rm accr}}\equiv
\mbh/\dot{m}_{{\rm BH}}$ is, for Bondi-Hoyle accretion
(Eq.~\ref{eq:bhl}), proportional to $\mbh^{-1}$. Therefore, depending
upon choices for various parameters, the initial growth of BHs can
proceed in one of two different ways. Firstly, in the regime where the
time scales over which gas accretion operates are very long, BHs may
grow primarily by mergers with other (seed mass) BHs until $m_{{\rm
    BH}}$ is large enough for the accretion rate to become
appreciable.  In this regime black holes initially grow at a rate
governed by the integrated mass of seed BHs they collide with.
Secondly, seed mass BHs may have accretion rates large enough for the
BHs to experience runaway growth until their accretion rate is limited
by feedback processes.

\begin{figure}
\begin{center}
\includegraphics[width=8.3cm,clip]{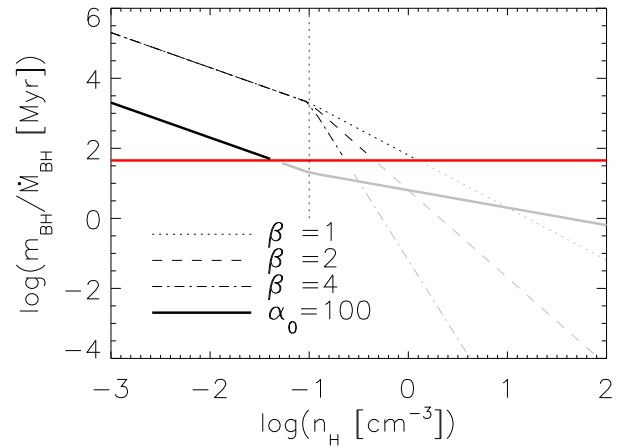}
\end{center}
\caption{BH growth times ($\mbh/\dot{m}_{{\rm BH}}$) as a function of
  the ambient gas density under various accretion models, all for a BH
  mass of $10^6\,{\rm M}_\odot$. The normalization of all black lines
  scales as $\mbh^{-1}$.  Lines are shown for both a constant-$\alpha$
  accretion model (solid, black/grey line) and for constant-$\beta$
  accretion models (all other black/grey lines).  The solid, red line
  shows the Salpeter time (the growth time for a BH accreting at the
  Eddington rate), and represents the lower limit on the BH growth
  time in the simulations. The grey section of each line represents
  the region where the accretion rate is greater than the Eddington
  rate. The vertical dotted line shows the star formation density
  threshold, $n_{\rm H}^*=10^{-1}$~cm$^{-3}$.  Above this density the
  gas follows the effective equation of state defined by
  Eq.~\ref{eq:effeos}.  For lower densities we have assumed the gas to
  be isothermal (only in this figure, not in the simulations). Note
  that the constant-$\alpha$ accretion model predicts that a
  $10^6\,{\rm M}_\odot$ BH will be growing at an Eddington limited
  rate even in gas with density at the star-formation threshold
  (0.1~cm$^{-3}$).}
\label{fig:bhga}
\end{figure}

\begin{figure}
\begin{center}
\includegraphics[width=8.3cm,clip]{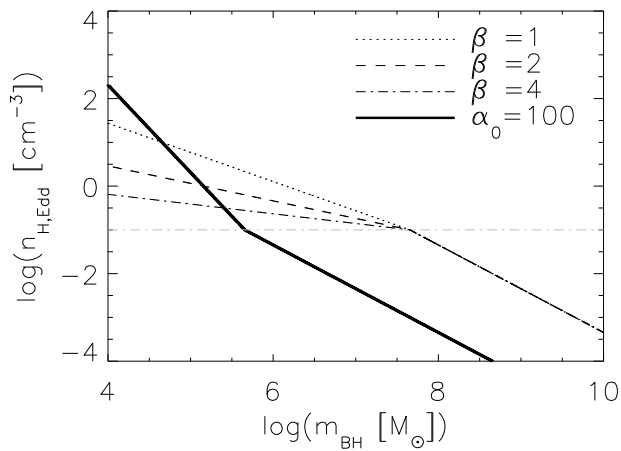}
\end{center}
\caption{The gas density above which the accretion rate onto a black
  hole becomes Eddington limited as a function of BH mass, assuming
  that gas with density above the critical density for star formation,
  $n_{\rm H}^*=10^{-1}$~cm$^{-3}$ has properties governed by the
  effective EOS and that gas below the critical density follows an
  isothermal EOS.  The thick, black line shows the behaviour of a
  constant-$\alpha$ accretion model, and all other lines show how
  models with a constant-$\beta$ accretion rate behave.  The grey line
  shows the critical density for star formation in our simulations.
  Except for very low BH masses, the constant-$\alpha$ model becomes
  Eddington limited at much lower gas densities than the
  constant-$\beta$ models.}
\label{fig:rhoedd}
\end{figure}

We can estimate the growth rate of BHs by noting that in our star
formation model we impose an effective equation of state on
star-forming gas, and as such can immediately calculate the local ISM
pressure, $P$, (and hence $c_s=\sqrt{\gamma P/\rho}$) from
\begin{equation}
P= P_{{\rm crit}}\Big(\frac{n_{{\rm H}}}{n_{{\rm H}}^*}\Big)^{\gamma_{{\rm eff}}}\,,
\label{eq:effeos}
\end{equation}
where $n_{{\rm H}}^*=0.1\,{\rm cm}^{-3}$ and $P_{{\rm crit}}$ are the
critical threshold density and pressure for star formation
respectively (see Sec.~\ref{sec:method}).  Fig.~\ref{fig:bhga} shows
BH growth times as a function of the ambient gas density for both
constant-$\alpha$ and constant-$\beta$ models, assuming here that the
BH is of mass $10^6\,{\rm M}_\odot$.  For the purposes of this plot we
assume that when gas densities are below the star-formation threshold
the EOS is isothermal, but note that in the simulations we calculate
the pressure self-consistently.  Following other authors, we set
$\alpha_0=100$ for the constant-$\alpha$ model. For the
constant-$\beta$ lines we set $\beta=[1,2,4]$.  The horizontal, red
line in this plot shows the growth time of a BH that is accreting at
the Eddington rate (i.e.\ the Salpeter time).  The Salpeter time
depends only upon physical constants and the BH radiative efficiency,
such that
\begin{equation}
t_{{\rm Salpeter}}\equiv \frac{\mbh}{\dot{m}_{{\rm Edd}}}=\frac{\epsilon_{\rm r}\sigma_{\rm T}c}{4\pi G m_{\rm p}}=4.5\times10^{5}\Big(\frac{\epsilon_{\rm r}}{0.1}\Big)\,{\rm yr}\,.
\end{equation}
It is immediately clear from Fig.~\ref{fig:bhga} that the choice of
accretion model strongly affects the local density at which the BH
growth becomes Eddington limited, with black holes accreting in the
constant-$\alpha$ model becoming Eddington limited at densities 1-2
orders of magnitude lower than the same black hole accreting in the
constant-$\beta$ model.

From the simulations we find that, for our chosen value of $m_{{\rm
    halo,min}}$ (=$100\,m_{{\rm DM}}$; see Sec.~\ref{sec:bhdem}) and
at high redshift, typical birth densities of BHs are $\sim 10-100$
times the star formation threshold.  Hence, in the regimes of
interest, in constant-$\alpha$ models all BHs of mass $> 10^5\,{\rm
  M}_\odot$ grow initially at close to the Eddington
rate\footnote{Note that the choice of sub-grid ISM model is a
  significant source of additional uncertainty in the accretion rates.
  For example, the use of the SH03 sub-grid multi-phase ISM can give
  rise to differences of almost an order of magnitude in the accretion
  rates of seed mass black holes due to the use of different effective
  equations of state.} (for a seed mass of $10^5\,{\rm M}_\odot$ and
typical initial gas densities of $10^1-10^2$~cm$^{-3}$ the initial
Eddington ratio is 0.037-0.37) until feedback effects reduce the local
gas density to values below the threshold for star formation.  From
Fig.~\ref{fig:bhga} we can see that for a BH of mass $10^6\,{\rm
  M}_\odot$ the accretion rate remains Eddington limited until the
local gas density falls to $n_{{\rm H}} \la 10^{-2}$~cm$^{-3}$.
Fig.~\ref{fig:rhoedd} presents this information in a slightly
different manner by showing the density above which a BH's accretion
rate is Eddington limited as a function of BH mass for the same
accretion models as in Fig.~\ref{fig:bhga}.  Again, it is clear that
the gas density below which the accretion rate depends on the density,
and can thus more easily be regulated by feedback from the AGN, has a
strong dependence on the accretion model used.  Take, for example, the
case of a BH of mass $10^8\,{\rm M}_{{\rm \odot}}$, here in a
constant-$\alpha$ model this BH will grow at the Eddington rate until
it can modulate its local density to below $10^{-3.5}$~cm$^{-3}$, more
than two orders of magnitude below the star-formation threshold.  This
is well within the regime where Bondi-Hoyle accretion is resolved in
the simulations.

In the constant-$\alpha$ model with $\alpha_0=100$ the growth of BHs
therefore proceeds as follows: Seed mass BHs (typical seed masses are
in the range $10^3-10^5{\rm M}_\odot$ in our simulations) grow
exponentially by Eddington limited accretion until feedback from the
BH has decreased the local ISM density to the point that growth is no
longer Eddington limited, and further energy output from the AGN can
decrease the accretion rate.  For BHs with masses greater than
$10^6\,{\rm M}_\odot$ self-regulation can only occur at densities
orders of magnitude below the star formation threshold
(Fig.~\ref{fig:rhoedd}).  In this regime we resolve Bondi-Hoyle
accretion, invalidating the assumption used to justify large values of
$\alpha$ in the first place.

In contrast, simulations that employ a constant-$\beta$ model, are not
necessarily Eddington limited from birth.  Taking again the case of a
$10^6\,{\rm {\rm M}}_\odot$ BH, and our fiducial value of $\beta=2$,
we see that the BH can decrease its accretion rate at a much higher
gas density, and as such the period of Eddington limited growth will
be much shorter than for the constant-$\alpha$ model.  The difference
in the gas density below which AGN accretion rates are no longer
Eddington limited can lead to large differences in the properties of
low mass galaxies and BH growth in small haloes.

\subsection{Efficient thermal feedback}
\label{sec:met-therm}
As discussed in Sec.~\ref{sec:met-feed}, it is necessary to impose a
minimum heating temperature in order to prevent BHs from
heating their surroundings before they have generated enough energy
for thermal feedback to become efficient.

Two parameters control how efficient this feedback may be: the number
of neighbours to heat ($\nheat$) and the temperature to which the
neighbours are heated ($\Delta T_{{\rm min}}$).  Two competing effects
control our choices for these parameters.  If $\Delta T_{{\rm min}}$
is too low then AGN heated gas will retain a low temperature and
therefore also a short cooling time (an analogous problem to the \lq
overcooling\rq\, of supernova heated gas in early cosmological
simulations \citep{katz96}, which is, however, usually attributed to
an overestimate of the gas density).  In this regime the energy will
be immediately radiated away, making AGN feedback ineffective.
Conversely, if $\Delta T_{{\rm min}}$ or $n_{{\rm heat}}$ are set too
high then the time scale over which BHs accrete enough energy to heat
$\nheat$ of their neighbours by an amount $\Delta T_{{\rm min}}$ will
become longer than either the dynamical time in the vicinity of the BH
(or the Salpeter time for Eddington limited growth), leading to
spurious growth as BHs are unable to self-regulate.

The choice of the minimum heating temperature, $\Delta T_{{\rm min}}$,
is motivated by the fact that we wish to choose the minimum value (and
so minimum time between heating events) for which the cooling time of
the heated gas is long enough that the energy is not immediately
radiated away.  In practice it was found that $\Delta T_{{\rm
    min}}=10^{8}$~K is the minimum temperature for which BH feedback
has a sufficient effect on galaxy clusters.  We return to this point
in Sec.~\ref{sec:res-pareffects}.

\begin{figure*}
\begin{center}
\includegraphics[width=\textwidth,clip]{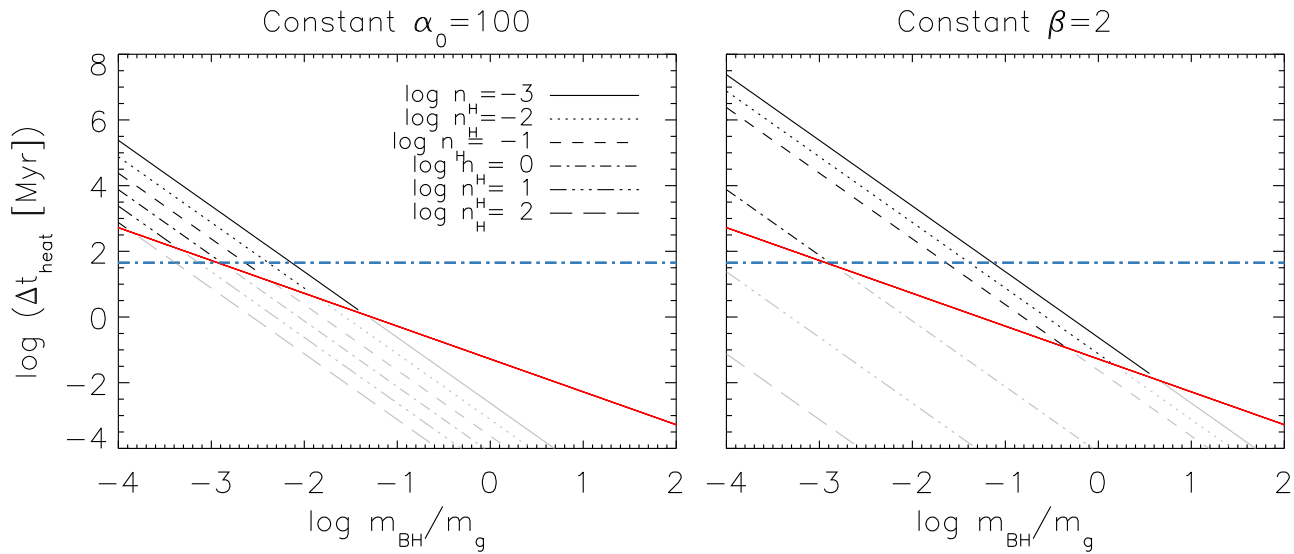}
\end{center}
\caption{Time between AGN heating events for Bondi-Hoyle (black and
  grey lines) and Eddington-limited (red lines) accretion as a
  function of the ratio between the BH mass and the simulation gas
  particle mass. The grey section of each line represents the region
  where $\dot{m}>\dot{m}_{{\rm Edd}}$.  In this plot we assume that
  $\nheat=1$ and $\Delta T_{{\rm min}}=10^8$~K.  All lines may be
  shifted vertically in proportion with $\nheat\Delta T_{{\rm min}}$.
  The horizontal blue line shows the Salpeter time ($m_{{\rm
      BH}}/\dot{m}_{{\rm Edd}}$). \emph{Left panel:} A
  constant-$\alpha$ accretion model with $\alpha_0=100$.  \emph{Right
    panel:} A constant-$\beta$ accretion model with $\beta=2$.  The
  spacings between the lines change when $n_{{\rm H}}>n_{{\rm H}}*$,
  because the accretion efficiency becomes proportional to $n_{{\rm
      H}}^\beta$.  For any Eddington limited accretion and BH masses
  greater than $10^{-3}\,\mg$ the time between heating events is
  shorter than the Salpeter time, enabling the BHs to self-regulate.}
\label{fig:bhg}
\end{figure*}

The second parameter, $\nheat$, is calibrated by noting that although
ideally we would like to allow AGN to heat gas instantaneously, our
finite resolution forces us to store energy until the feedback is
effective, hence introducing a delay to AGN heating.  We can minimise
the effect of this delay by noting that the numerically imposed time
between heating events should be lower than the typical time-scales of
dynamical processes that affect AGN feedback.  If $\nheat$ is set too
high then it is possible that the amount of time taken for a BH to
accrete enough energy to perform a heating event would be large enough
that we see spurious growth.  We can quantify this effect by
calculating the mean time between heating events for BHs of different
masses in different density environments.  This is demonstrated in
Fig.~\ref{fig:bhg}, which shows the mean time between heating events
as a function of $\mbh/\mg$ for models with both constant-$\beta$ and
constant-$\alpha$ accretion rates.  Plotting the heating time as a
function of this mass ratio allows us to make this plot in a
resolution independent manner.  Fig.~\ref{fig:bhg} assumes $\Delta
T_{{\rm min}}=10^8$~K and $\nheat=1$, but all lines can be shifted
vertically in proportion with the quantity $\Delta T_{{\rm
    min}}\nheat$.  To make the time between feedback events as small
as possible we should choose $n_{{\rm heat}}$ as small as possible,
but it is not immediately obvious that BHs will be able to regulate
their growth if we heat only a small number of neighbours of an AGN
that it will be able to self-regulate.  However, we found from
numerical tests (Sec.~\ref{sec:res-pareffects}) that $n_{{\rm
    heat}}=1$ is sufficient for BH feedback to be effective, and so
$n_{{\rm heat}}=1$ is the parameter value used in our fiducial
simulations.  The dependence of our results on the two parameters
$\Delta T_{{\rm min}}$ and $n_{{\rm heat}}$ will be discussed in
Sec.~\ref{sec:res-pareffects}.

The next model parameter is $\epsilonf$, the efficiency with which
energy radiated from the BH is coupled to the ISM. The parameter
$\epsilonf$ sets the normalizations of the global BH density and the
BH-galaxy scaling relations.  We therefore tune $\epsilonf$ after
setting all of the other parameters in order to match the redshift
zero observations (Sec.~\ref{sec:res-pareffects},
Fig.~\ref{fig:params_bhdens}d) and find that a value of
$\epsilon_{{\rm f}}=0.15$ provides a good match to the observations.

\subsection{Black hole seed mass and minimum halo mass}
\label{sec:bhdem}

Our initial choice for the halo mass into which we insert a BH is
motivated by the fact that we wish for every resolved halo with
$m_{{\rm halo}} \gg m_{{\rm seed}}$ to contain a seed BH.  We
therefore choose to place BH seeds into haloes of a constant particle
number.  Using $m_{{\rm halo,min}}=100\,m_{{\rm DM}}$ ensures that
haloes containing BHs are always well defined \citep[e.g.][]{diem07}.
The choice of a constant particle number halo mass also has the
advantage that if we change the simulation mass resolution, BHs will
still be placed into the smallest allowable mass of dark matter haloes
without the need to tune any parameters. Note, however, that this
prescription will have to be changed for simulations that have
sufficient resolution for $100\,m_{{\rm DM}}$ to be comparable or
smaller than the minimum halo mass expected to be able to form seed
mass BHs.

\begin{figure}
\begin{center}
\includegraphics[width=8.3cm,clip]{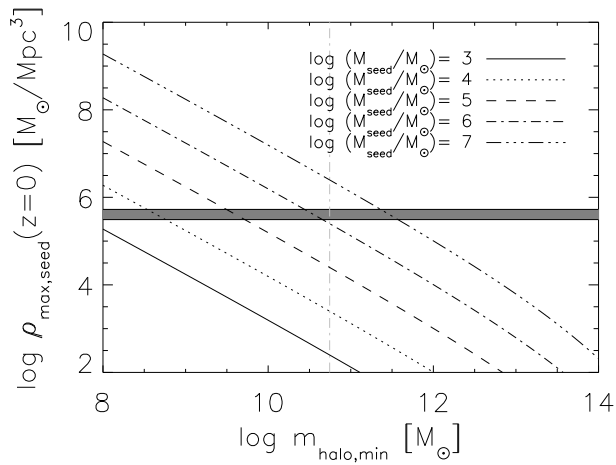}
\end{center}
\caption{Maximum possible contribution to the cosmic BH density from
  seed mass BHs, as a function of the minimum dark matter halo mass,
  $m_{{\rm halo,min}}$, assuming the dark matter mass function of Reed
  et al. (2006). Each black line corresponds to a different BH seed
  mass, as indicated in the legend.  The horizontal, grey shaded
  region shows the observed cosmic black hole density at redshift
  $z=0$ (Shankar et al. 2004), and the vertical line indicates
  $m_{{\rm halo,min}}$ for our fiducial simulation, which has a DM
  mass resolution of $8.64\times 1-^7\,{\rm M}_\odot$.  For our
  fiducial BH seed mass of $m_{{\rm seed}}=1.2\times1-^5\,{\rm
    M}_\odot$ we see that the maximum possible contribution of seed
  mass BHs to the global BH density is much lower than the $z=0$
  observations.}
\label{fig:seeddens}
\end{figure}

Given the minimum halo mass into which we place BH seeds, we must
ensure that the integrated number of seed BHs generated between
redshifts $z=\infty$ and zero is much smaller than the observed cosmic
BH density.  We can obtain an upper limit on the cumulative cosmic
density of BH seeds by taking the redshift zero dark matter halo mass
function $f(m)=n(m)dm$ assuming that \emph{all} collapsed mass was
assembled through mergers of critical mass haloes:
\begin{equation}
\rho_{{\rm seed}}(m_{{\rm halo,min}})<\frac{m_{{\rm seed}}}{m_{{\rm halo,min}}}\int^\infty_{m_{{\rm halo,min}}}mf(m)dm\,.
\end{equation}
This quantity is plotted for a number of values of the seed BH mass in
Fig.~\ref{fig:seeddens}, where the two vertical grey lines represent
the masses of haloes of 100 DM particles (=$m_{{\rm halo,min}}$) in
our fiducial simulations run at the mass resolution as the OWLS
  runs of Schaye et al.~(in preparation).

Given choices for $\Delta T_{{\rm min}}$ and $\nheat$, we can use
Fig.~\ref{fig:bhg} to place a minimum limit on the BH seed mass,
$m_{{\rm seed}}$.  Here, we show the time between heating events as a
function of BH mass, for both constant-$\alpha$ and constant-$\beta$
models.  The time between heating events for BHs accreting at the
Eddington accretion rate are shown as red lines in each panel.  We now
note that we require BH heating to occur regularly in high density
environments.  In particular, in order for a BH to be able to
effectively self-regulate its own growth, we require that the
numerically imposed minimum duty-cycle, $\Delta t_{{\rm heat}}$, is
less than the Salpeter time, the characteristic growth time for black
holes accreting at the Eddington rate.  It is clear from
Fig.~\ref{fig:bhg} by comparing the BH Salpeter time (blue line) to
the BH duty cycle that in high density environments ($n_{{\rm
    H}}>10^2\,{\rm cm}^{-3}$) this condition is satisfied only if the
BH mass is greater than $10^{-3}\mg$.  This provides a minimum allowed
seed mass in our models.

However, in addition to ensuring that BHs grow in a physical manner
and that their feedback can be effective, we must also satisfy various
observational constraints.  Most fundamentally, it is known that the
present day cosmic BH density is $(4.2\pm 1.1)\times
10^5$~M$_{\odot}$/Mpc$^3$ \citep{shan04} or
$(4.22^{+1.75}_{-1.22})\times 10^5$~M$_{\odot}$/Mpc$^3$
\citep{marc04}, although we caution that a more accurate consideration
of the effects of cosmology may lead to a slightly higher
determination of the BH density \citep{grah07b}.  In order that we do
not violate these observational constraints in the presence of
substantial BH growth through accretion, we require the $z=0$ global
seed density to be much smaller than the observed BH mass density.  We
now ensure that -- given all of our other parameter choices --
$m_{{\rm seed}}=10^{-3}\,\mg$ does not violate this constraint on the
global BH density.  For our simulations $10^{-3}\,\mg$ corresponds to
a BH seed mass of $1.2\times10^5\,{\rm M}_{\odot}$. It is clear from
Fig.~\ref{fig:seeddens} that the maximum possible contribution of the
seed BH mass to the cosmic density is at least a factor of 10 less
than the redshift zero observations, which we indicate by the grey,
horizontal shaded region.  We will see in Fig.~\ref{fig:params_bhdens}
that, as expected, the actual contributions of seed BHs to the total
cosmic density are much smaller than this value.

An additional observational constraint is placed by the well-defined
relation between the mass of a BH and the mass of the bulge component
of a galaxy, $\mbh\approx 0.006\,m_{{\rm bulge}}$ \citep{magg98}.  In
simulations it is more convenient to work with the relation between BH
mass and dark matter halo mass, investigated by \cite{ferr02}, who
found that in halos of $10^{12}\,{\rm M}_{\odot}$ the ratio $m_{{\rm
    BH}}/m_{{\rm halo}}\sim 10^{-5}$.  This provides a second, related
constraint on the mass of seed BHs: we wish to place them below this
relationship so that they can subsequently grow on to the observed
redshift zero relation\footnote{In some BH seed generation scenarios,
  for example the direct collapse of matter in haloes we expect BH
  seeds to reside initially above the observed scaling relations. We
  will show in Sec. \ref{sec:seedmodel} that in our models BHs grow
  onto the BH scaling relations regardless of whether they are
  initially placed above or below the relations.}.  For our parameter
choices ($m_{{\rm seed}}=10^{-3}\,\mg$ and $m_{{\rm
    halo,min}}=100\,m_{{\rm DM}}$)
\begin{equation}
\frac{m_{{\rm seed}}}{m_{{\rm halo,min}}}=\frac{10^{-3}}{100}\Big(\frac{\Omega_b}{\Omega_m-\Omega_b}\Big)=2.1\times10^{-6}\,,
\end{equation}
where the last equality assumes our chosen cosmology.  This ratio is
indeed much smaller than the observed value.

Finally, we note that our fiducial choice of seed mass, $m_{{\rm
    seed}}=10^{-3}\,\mg$, will need to be modified for simulations
that have sufficient resolution for this value to be below the
expected BH seed masses.

\subsection{Comparison with previous work}
\label{sec:met-pars}

Through the arguments in the previous sections we were able to specify
values for all of our model parameters. These fiducial parameter
values are summarised in Table~\ref{tab:parlist}.  The AGN model
developed in this paper is a modification of that introduced in S05,
and used thereafter in a large number of works.  As such it is
instructive to compare our parameter choices with those employed in
other studies, as collected in Table~\ref{tab:allparlist}.

We turn our attention first to the AGN feedback efficiency,
$\epsilonf$.  The value used in the present study ($\epsilonf=0.15$)
is significantly higher than that used in previous published studies,
which all assume $\epsilon_{{\rm f}}=0.05-0.1$.  We can account for
this difference if we note that, unlike the other studies, we do not
employ the SH03 subgrid model for the ISM.  Use of a different subgrid
model for the unresolved ISM is likely to lead to differences in the
amount of radiative losses, as the effective density and temperature
of the ISM differ significantly between the two models.

We note that apart from differences in the ISM model and AGN heating
mechanisms, the strength of the AGN feedback depends only on the
parameter combination $\epsilonf\epsilon_{{\rm r}}$, and that there is
significant leeway in the value of $\epsilon_{{\rm r}}$.  All studies
presented in Table \ref{tab:allparlist} assume $\epsilon_{{\rm
    r}}=10\%$, but values close to $\epsilon_{{\rm r}}=20\%$ are
possible for thin-disc accretion on to a Kerr BH \citep{yu02,thor74}.
Recent observational determinations of $\epsilon_{\rm r}$ span the
full range of allowable values: $\epsilon_{{\rm r}}=30-35\%$
\citep{wang06}, $\epsilon_{{\rm r}}=15\%$ \citep{elvi02,yu08},
$\epsilon_{{\rm r}}=7-8\%$ \citep{cao08,mart08} depending upon the
specific assumptions and models used in each study.

The ratio of the minimum halo mass to the seed mass is similar in all
of the cosmological studies, with the exception of the work of
\cite{khal08}, who performed zoomed cosmological simulations of an
individual object.  In order to avoid numerical issues when
$\mbh<\mg$, these authors forced the BH to accrete very
quickly at early times before artificially halting its accretion
until the stellar mass of the halo becomes large enough that the BH
lies on the observed $\mbh-m_{{\rm *}}$ relation.

We see also from Table \ref{tab:allparlist} that the minimum halo mass
($m_{{\rm halo, min}}$) is consistent between all of the published
cosmological studies to within a factor of 5 ($1-5\times10^{10}\,{\rm
  M}_\odot$).  We will show in Sec.~\ref{sec:res-pareffects} that, for
the constant-$\alpha$ accretion model assumed in previous studies, AGN
feedback affects all haloes into which seed mass black holes are
placed, and that changing $m_{{\rm halo,min}}$ by a factor of of ten
has a large effect on the BH properties of the less massive haloes in
the simulation. Care should therefore be taken when comparing results
of different simulations.

The two areas where our work differs most from previous models are the
BH accretion model and the SN feedback model.  We turn our attention
first to the SN model and note that almost all previous studies employ
the work of SH03, whereas we use the method of \citet{dall08}.
Contrary to SH03, SN winds in our model are local and not
hydrodynamically decoupled from the surrounding gas.  Hydrodynamical
decoupling of supernova heated gas, as implemented in the SH03 model,
guarantees that when gas is kicked it is able to escape the ISM
(although it may subsequently return). On the other hand, if the
hydrodynamic forces are taken into account, the gas can remain
confined by pressure forces, and if it does manage to escape it may
drag along much of its neighbouring gas.  It was shown in
\cite{dall08} that in high-resolution simulations of individual disc
galaxies this change fundamentally alters the structure of the
galactic disc and that hydrodynamically coupled winds and generate
galactic outflows with properties broadly comparable with
observations.  We demonstrate in a companion work that SN feedback has
a large effect on the properties of the AGN population and, as such,
it is important to investigate a number of SN feedback prescriptions.

The second area where our model differs significantly from others in
the literature is in the accretion model.  In the nomenclature of this
paper, all of the previous AGN models are constant-$\alpha$ accretion
models in which $\alpha_0=100-300$.  We show in later sections that
the accretion model represents one of the most crucial elements of an
AGN model, and that all results are very sensitive to the way in which
BHs are allowed to accrete.

We show in this work that the results derived from simulations depend
on aspects of the BH model that are homogeneous between the studies
that have thus far been published.  The present work -- which is
carried out using different techniques and parametrizations for much
of the sub-grid modelling -- therefore provides a way to investigate
the robustness of the models.

\begin{table*}\begin{center}
\caption{Model parameters for AGN feedback studies in the literature.
  Columns are as follows: 1) Reference; 2) Efficiency with which
  energy emitted by a BH is coupled into the ambient gas; 3) Radiative
  efficiency of BH accretion discs; 4) Multiplication factor for the
  Bondi-Hoyle accretion rate; 5)For star forming gas the Bondi-Hoyle
  accretion rate is multiplied by $(n_{{\rm H}}/10^{-1}\,{\rm
    cm}^{-3})^\beta$ (Eq.~\ref{eq:beta}); 6) Mass of seed BHs; 7)
  Minimum halo mass in which black hole seeds are placed; 8) Number of
  DM particles corresponding to a halo of mass $m_{{\rm halo,min}}$,
  ranges indicate the difference between the highest and lowest
  resolution simulations used in each study; 9) The ratio of the BH
  seed mass to the minimum halo mass; 10) Type of simulation: Iso.,
  isolated model disc galaxy. Zoom, zoomed simulation of individual
  object in a cosmological volume.  Cosmo., uniform mass resolution
  simulation of a cosmological volume; 11) Reference for star
  formation model used in the study: Springel \& Hernquist (2003)
  (SH03), Schaye \& Dalla Vecchia (2008) (SD08); 12) Reference for
  supernova wind model used in this study: Springel \& Hernquist
  (2003) (SH03), Dalla Vecchia \& Schaye (2008) (DS08)}
\begin{tabular}{l|l|l|l|l|l|l|l|l|l|l|l}
Study  & $\epsilonf$ & $\epsilon_{{\rm r}}$ & $\alpha_0$ & $\beta$ & $\frac{m_{{\rm seed}}}{{\rm M}_{\odot}}$ & $\frac{m_{{\rm halo,min}}}{M_{\odot}}$ & $N_{{\rm halo,min}}$ & $\frac{m_{{\rm seed}}}{m_{{\rm halo,min}}}$ & Type & SF Model & Wind Model\\
(1) & (2) & (3) & (4) & (5) & (6) & (7)& (8) & (9) & (10) & (11) & (12)\\
\hline
Springel et al. (2005)     & 0.05 & 0.1 & 100  &  0 & $10^5$                 & n/a                    & n/a      & n/a  & Iso.                      & SH03 & SH03\\
Robertson et al. (2006)    & 0.05 & 0.1 & 100  &  0 & $10^5$                 & n/a                    & n/a      & n/a  & Iso.                      & SH03 & SH03\\
Sijacki et al. (2007)      & 0.05 & 0.1 & 100  &  0 & $10^5$                 & $5\times 10^{10}$       & 2260     & $2\times 10^{-6}$  & Zoom         & SH03 & SH03\\
Sijacki et al. (2007)      & 0.05 & 0.1 & 100  &  0 & $10^5$                 & $5\times 10^{10}$       & 285-606  & $2\times 10^{-6}$  & Cosmo.       & SH03 & SH03\\
Johansson et al. (2008)    & 0.05 & 0.1 & 100  &  0 & $10^5$                 & n/a                    & n/a      & n/a               & Iso.         & SH03 & SH03\\
Di Matteo et al. (2008)    & 0.05 & 0.1 & 100  &  0 & $10^5$                 & $1\times 10^{10}$       & 36-363   & $1\times 10^{-5}$  & Cosmo.       & SH03 & SH03\\
Khalatyan et al. (2008)    & 0.1  & 0.1 & 300  &  0 & n/a                    & n/a                    & 1463      & n/a              & Zoom        & SH03 & SH03\\
This study                 & 0.15  & 0.1 &  1    &  2 & $10^{-3}m_{\rm g}$      & 100$m_{{\rm DM}}$        & 100      & $2\times 10^{-6}$  & Cosmo.        & SD08 & DS08\\

\end{tabular}
\label{tab:allparlist}
\end{center}\end{table*}

%
%
\section{Simulation results}
\label{sec:results}

\begin{figure*}
\begin{center}
\includegraphics[width=\textwidth,clip]{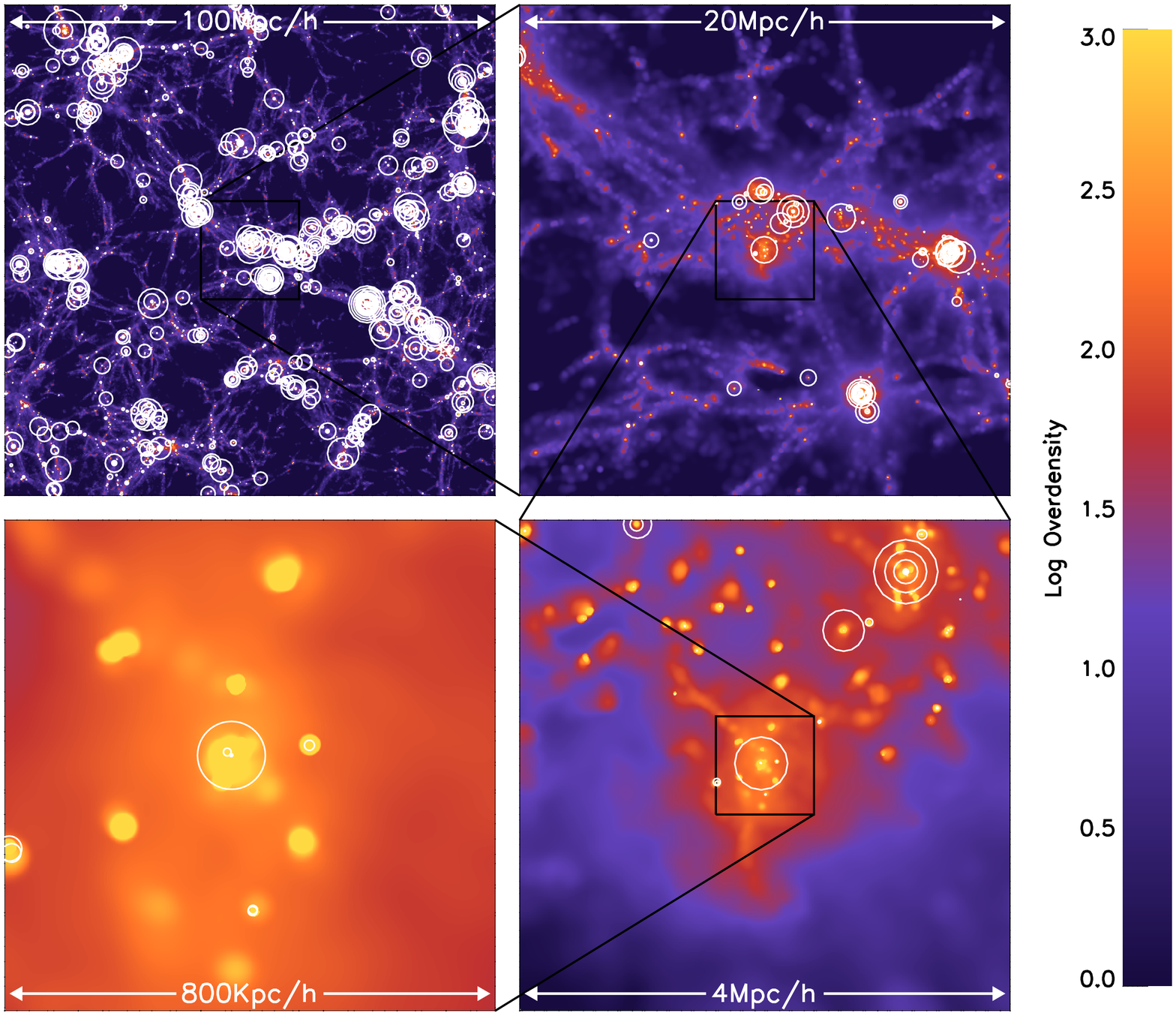}
\end{center}
\caption{Successively zoomed projections of the gas density in a
  3~Mpc/$h$ thick slice from our $L100N512$ simulation at redshift
  zero.  BHs are represented in this plot by open circles and the area
  of each circle is proportional to the logarithm of the BH mass.  The
  largest circle in the lower-left panel represents a BH of mass
  $3\times10^{7}\,{\rm M}_\odot$}
\label{fig:pretty}
\end{figure*}

In this section we first introduce the simulation set used in this
paper (Sec.~\ref{sec:simlist}) before demonstrating that the fiducial
simulations reproduce redshift zero observational results and
quantifying how robust our model is to poorly constrained parameter
choices (Sec.~\ref{sec:res-pareffects}).  To begin with, however, we
show for illustrative purposes the gas density in a 3~Mpc/$h$
thick slice from our 100~Mpc/$h$ ($L100N512$) simulation at redshift
zero (Fig.~\ref{fig:pretty}).  Each panel represents a successive
factor of five zoom.  The bottom-left panel shows a region 800~kpc/$h$
across, centered on a $3\times10^{7}\,{\rm M}_\odot$ BH, contained in
a halo with a stellar mass of $3\times 10^{10}\,{\rm M}_\odot$ and a
dark matter mass of $2\times 10^{12}\,{\rm M}_\odot$.  Circles in this
plot represent the locations of BHs with the area of each circle
proportional to the logarithm of the mass of the BH.

\subsection{Simulation list}
\label{sec:simlist}
In order to explore parameter space we have run a large number of
smaller simulations, the details of which are summarised in Table
\ref{tab:sims}.  Simulation names are of the form \emph{LxxxNyyy},
where \emph{xxx} represents the simulation box size in comoving
Mpc/$h$ and \emph{yyy} is the cube root of the initial number of dark
matter and gas particles.  For example, the simulation denoted
$L100N512$ refers to a comoving simulation volume of 100 Mpc/$h$,
which contains $512^3$ dark matter particles and an equal number of
baryonic particles. Simulations for which one of the parameters was
changed from its default value are denoted by appending a descriptive
suffix to the end of the simulation name. For example, simulations
without AGN feedback are named \emph{LxxxNyyyNOAGN}, and correspond to
the simulations denoted \emph{REF\_LxxxNyyy} in the OWLS project
(Schaye et al., in preparation).

%
%
\begin{table*}\begin{center}
\caption{Summary of simulation parameters. Columns are as follows: 1)
  Name of simulation; 2) Comoving box size; 3) Number of DM and gas
  particles; 4) Final redshift; 5) Multiplication factor for the
  Bondi-Hoyle accretion rate (Eq.~\ref{eq:bhl}); 6) For star forming
  gas the Bondi-Hoyle accretion rate is multiplied by $(n_{{\rm
      H}}/10^{-1}\,{\rm cm}^{-3})^\beta$ (Eq.~\ref{eq:beta}); 7)
  Fraction of radiated energy that is coupled back into the ISM
  (Eq.~\ref{eq:epsilon}); 8) Initial mass of sub-grid BH; 9) Minimum
  halo mass into which a seed BH is placed; 10) Minimum AGN heating
  temperature difference; 11) Number of particles heated per event.  Entries in
  boldface represent parameters that have been changed from the value
  they take in the default model. }
\begin{tabular}{l|r|l|l|r|l|l|l|r|l|r}
Identifier & $\frac{L}{{\rm Mpc}/h}$ & $N_{{\rm gas}}$ & $z_{{\rm end}}$& $\alpha_0$ & $\beta$ & $\epsilonf$ & $\frac{m_{{\rm seed}}}{\mg}$ & $\frac{m_{{\rm halo,min}}}{m_{{\rm DM}}}$ & $\frac{\Delta T_{{\rm heat}}}{10^8\,{\rm K}}$ & $\nheat$ \\
(1) & (2) & (3) & (4) & (5) & (6) & (7) & (8) & (9) & (10) & (11) \\
\hline
\multicolumn{11}{|c|}{Simulations for parameter studies} \\
\emph{L050N256}        & 50.0   & $256^3$ & 0 & 1 & 2 & 0.15 & 0.001 & 100 & 1 & 1 \\
\emph{L050N256VLOEPS}  & 50.0   & $256^3$ & 0 & 1 & 2 & {\bf 0.0375} & 0.001 & 100 & 1 & 1 \\
\emph{L050N256LOEPS}   & 50.0   & $256^3$ & 0 & 1 & 2 & {\bf 0.075} & 0.001 & 100 & 1 & 1 \\
\emph{L050N256HIEPS}   & 50.0   & $256^3$ & 0 & 1 & 2 & {\bf 0.3} & 0.001 & 100 & 1 & 1 \\
\emph{L050N256VHIEPS}  & 50.0   & $256^3$ & 0 & 1 & 2 & {\bf 0.6} & 0.001 & 100 & 1 & 1 \\
\emph{L050N256HIHALO}  & 50.0   & $256^3$ & 0 & 1 & 2 & 0.15 & 0.001 & {\bf 1000} & 1 & 1 \\
\emph{L050N256HISEED}  & 50.0   & $256^3$ & 0 & 1 & 2 & 0.15 & {\bf 0.01} & 100 & 1 & 1 \\
\emph{L050N256LOSEED}  & 50.0   & $256^3$ & 0 & 1 & 2 & 0.15 & {\bf 0.0001} & 100 & 1 & 1 \\
\emph{L050N256HINHEAT} & 50.0   & $256^3$ & 0 & 1 & 2 & 0.15 & 0.001 & 100 & 1 & {\bf 10} \\
\emph{L050N256B0}     & 50.0   & $256^3$ & 0 & 1 & {\bf 0} & 0.15 & 0.001 & 100 & 1 & 1 \\
\emph{L050N256B1}     & 50.0   & $256^3$ & 0 & 1 & {\bf 1} & 0.15 & 0.001 & 100 & 1 & 1 \\
\emph{L050N256B4}     & 50.0   & $256^3$ & 0 & 1 & {\bf 4} & 0.15 & 0.001 & 100 & 1 & 1 \\
\emph{L050N256A100B0}  & 50.0   & $256^3$ & 0 & {\bf 100} & {\bf 0} & 0.15 & 0.001 & 100 & 1 & 1 \\
\emph{L050N256A1000B0} & 50.0   & $256^3$ & 0 & {\bf 1000}& {\bf 0} & 0.15 & 0.001 & 100 & 1 & 1 \\
\emph{L050N256T7}      & 50.0   & $256^3$ & 0 & 1 & 2 & 0.15 & 0.001 & 100 & {\bf 0.1} & 1 \\
\emph{L050N256NOAGN}   & 50.0   & $256^3$ & 0 & -- & -- & -- & -- & -- & -- & -- \\
\multicolumn{11}{|c|}{Simulations for box size and resolution tests}\\
\emph{L100N128}        & 100.0  & $128^3$ & 0 & 1 & 2 & 0.15 & 0.001 & 100 & 1 & 1 \\
\emph{L100N128NOAGN}   & 100.0  & $128^3$ & 0 & -- & -- & -- & -- & -- & -- & -- \\
\emph{L100N256}        & 100.0  & $256^3$ & 0 & 1 & 2 & 0.15 & 0.001 & 100 & 1 & 1 \\
\emph{L100N256NOAGN}   & 100.0  & $256^3$ & 0 & -- & -- & -- & -- & -- & -- & -- \\
\emph{L100N512}        & 100.0  & $512^3$ & 0 & 1 & 2 & 0.15 & 0.001 & 100 & 1 & 1 \\
\emph{L100N512NOAGN}   & 100.0  & $512^3$ & 0 & -- & -- & -- & -- & -- & -- & -- \\
\emph{L025N128}        & 25.0   & $128^3$ & 0 & 1 & 2 & 0.15 & 0.001 & 100 & 1 & 1 \\
\label{tab:sims}
\end{tabular}
\end{center}\end{table*}

\subsection{The effects of changing the model parameters}
\label{sec:res-pareffects}

\begin{figure*}
\begin{center}
\includegraphics[width=0.8\textwidth,clip]{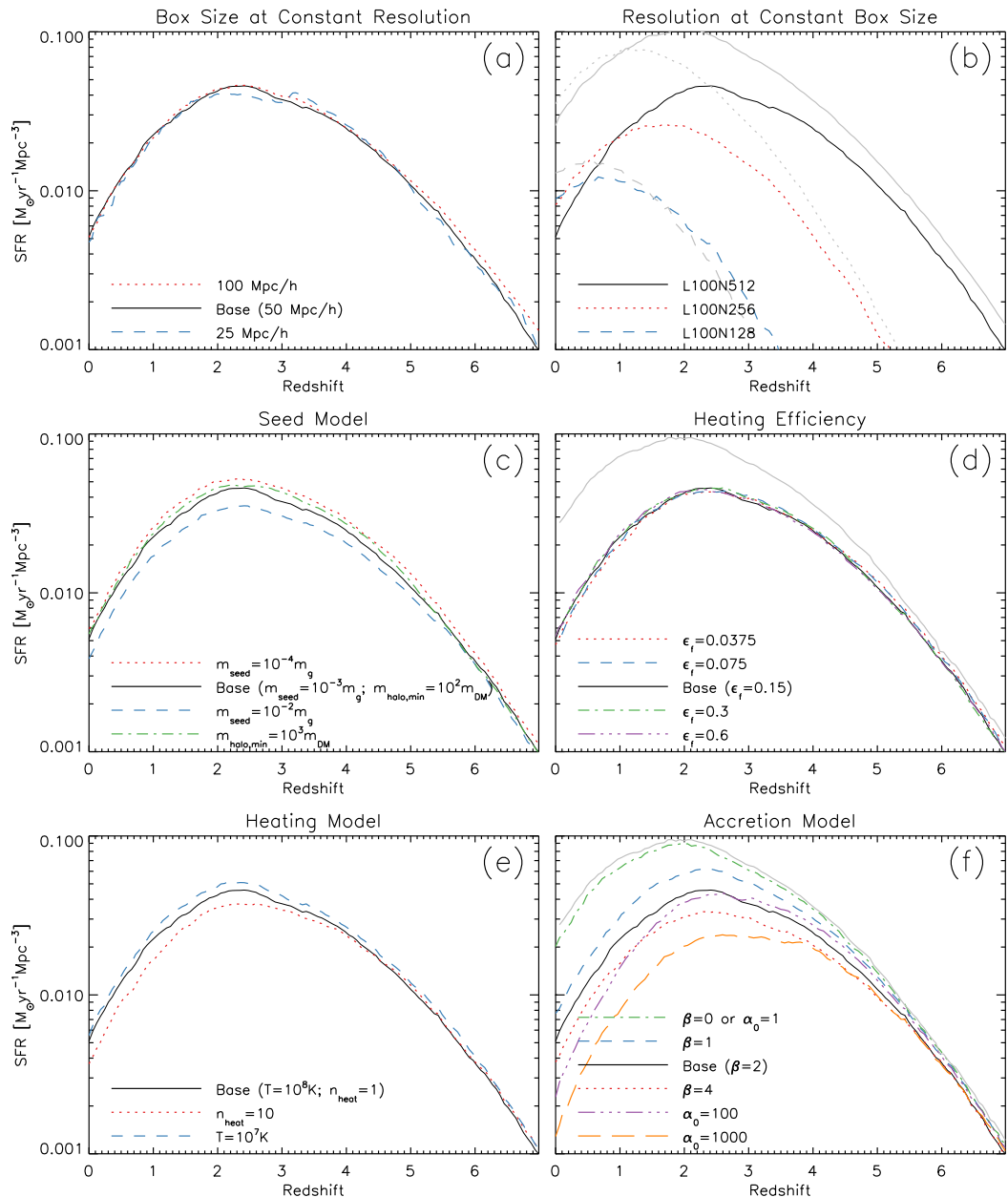}
\end{center}
\caption{The dependence of the global star formation history on
  various sets of parameters.  \emph{ (a)}: The effect of changing the
  simulation volume at a constant resolution; \emph{ (b)}:The effect
  of changing the resolution at a constant box size;
  \emph{(c)}:Simulations with different prescriptions for the
  generation of seed BHs ($m_{{\rm halo,min}}$, $m_{{\rm seed}}$) ;
  \emph{(d)}: Simulations with different values of the heating
  efficiency ($\epsilonf$);\emph{(e)}: The effect of changing the way
  in which AGN feedback energy is distributed to the surrounding gas
  particles ($\Delta T_{{\rm min}}$, $\nheat$); \emph{(f)}: The effect
  of changing the BH accretion model ($\alpha_0$, $\beta$).  Curves in
  grey represent simulations that do not include AGN feedback and the
  solid black curve in each panel represents the fiducial simulation
  (\emph{L050N256}).}
\label{fig:params_sfr}
\end{figure*}

\begin{figure*}
\begin{center}
\includegraphics[width=0.8\textwidth,clip]{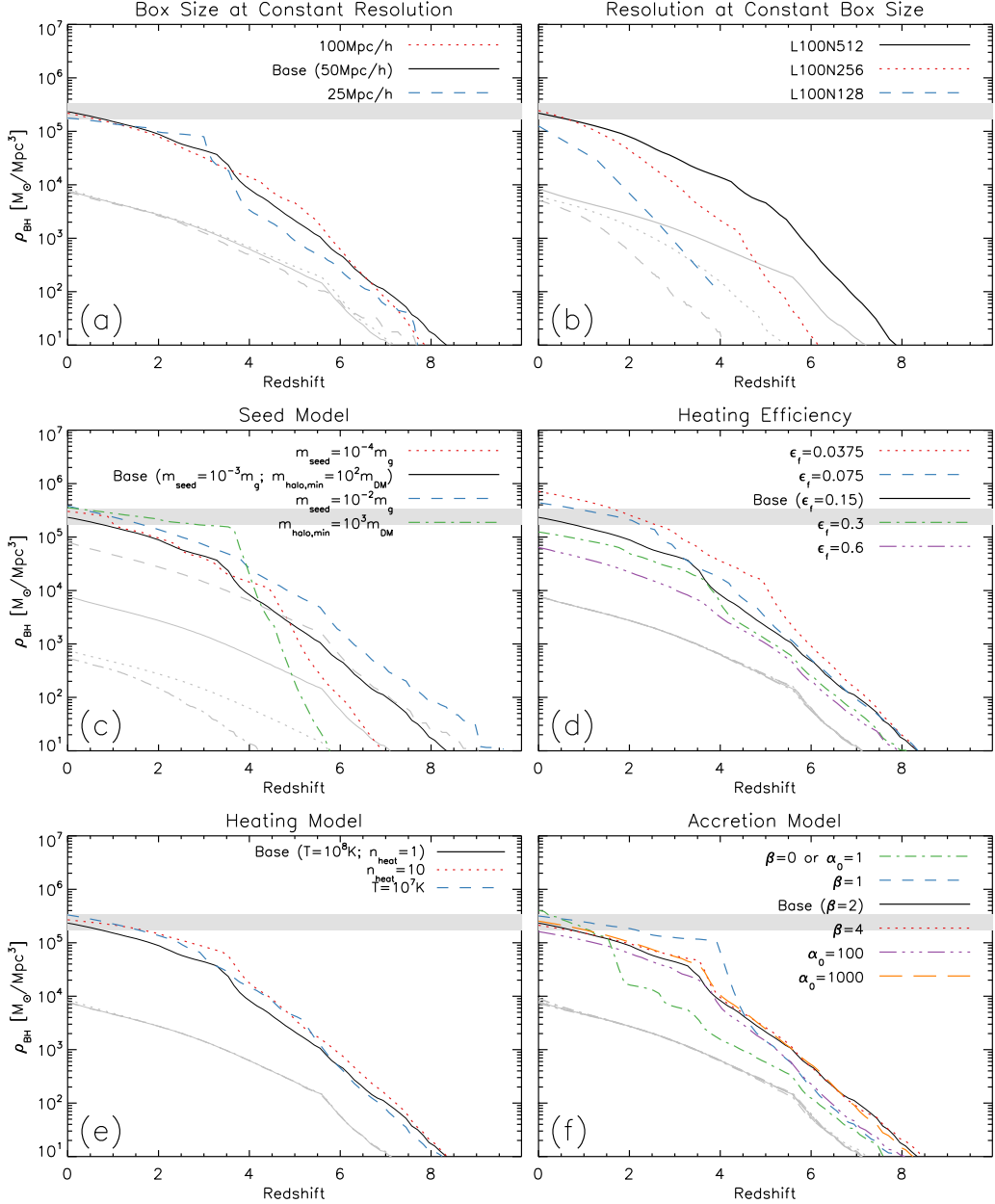}
\end{center} 
\caption{The dependence of the evolution of the the cosmic BH
  density on the parameters of the AGN model.  The panels are the same
  as in Fig.~\ref{fig:params_sfr}, the shaded grey area represents
  the redshift zero BH density as measured by Shankar et
  al. (2004).  The grey curves show the cumulative density in seed
  BHs.  The solid, black curve in each panel represents the
  fiducial simulation (\emph{L050N256}).}
\label{fig:params_bhdens}
\end{figure*}

\begin{figure*}
\begin{center}
\includegraphics[width=0.8\textwidth,clip]{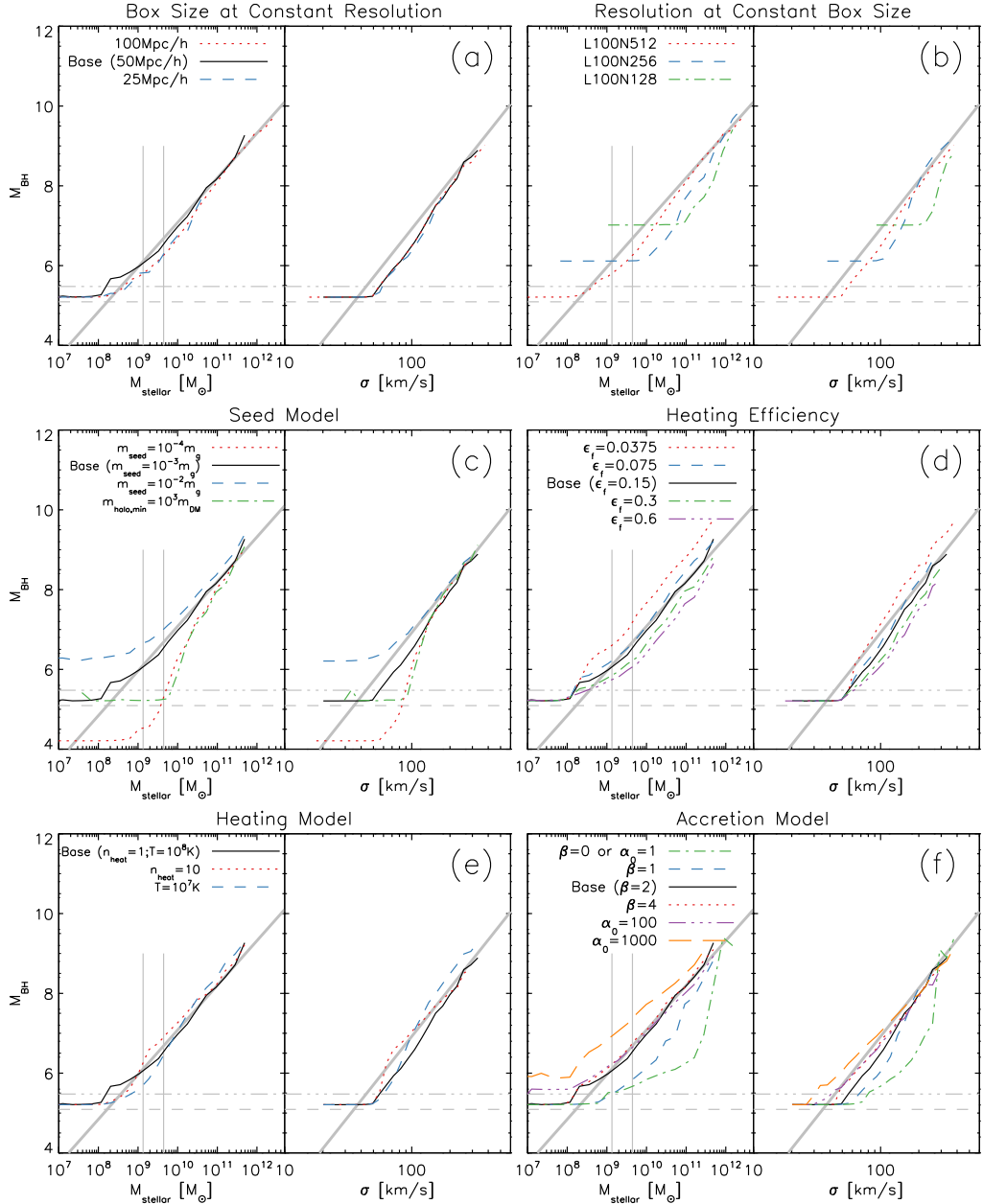}
\end{center}
\caption{The Magorrian relation and the $\mbh-\sigma$ relation for
  each set of simulations. Line styles and panels correspond to those
  in Fig.~\ref{fig:params_sfr}.  The simulated stellar velocity
  dispersion is calculated as the mean of the three one-dimensional
  velocity dispersions of the stellar particles in each halo.  The
  pale, solid, grey line in each plot represents the redshift zero
  observations of Haring \& Rix (2004) (left panel) and Tremaine et
  al. (2002) (right panel).  The solid, black curve in each panel
  represents the fiducial simulation (\emph{L050N256}).  The
  horizontal, triple-dot-dashed line indicates the mass that a BH
  requires to begin to heat its surroundings, and the horizontal grey,
  dashed line indicates the black hole seed mass in the fiducial
  simulation.  The two solid vertical lines show, for the fiducial
  simulation, the stellar mass above which 50\% and 90\% of haloes
  contain a BH massive enough to have performed at least one heating
  event.}
\label{fig:params_mbhmhalo}
\end{figure*}

\begin{figure*}
\begin{center}
\includegraphics[width=0.8\textwidth,clip]{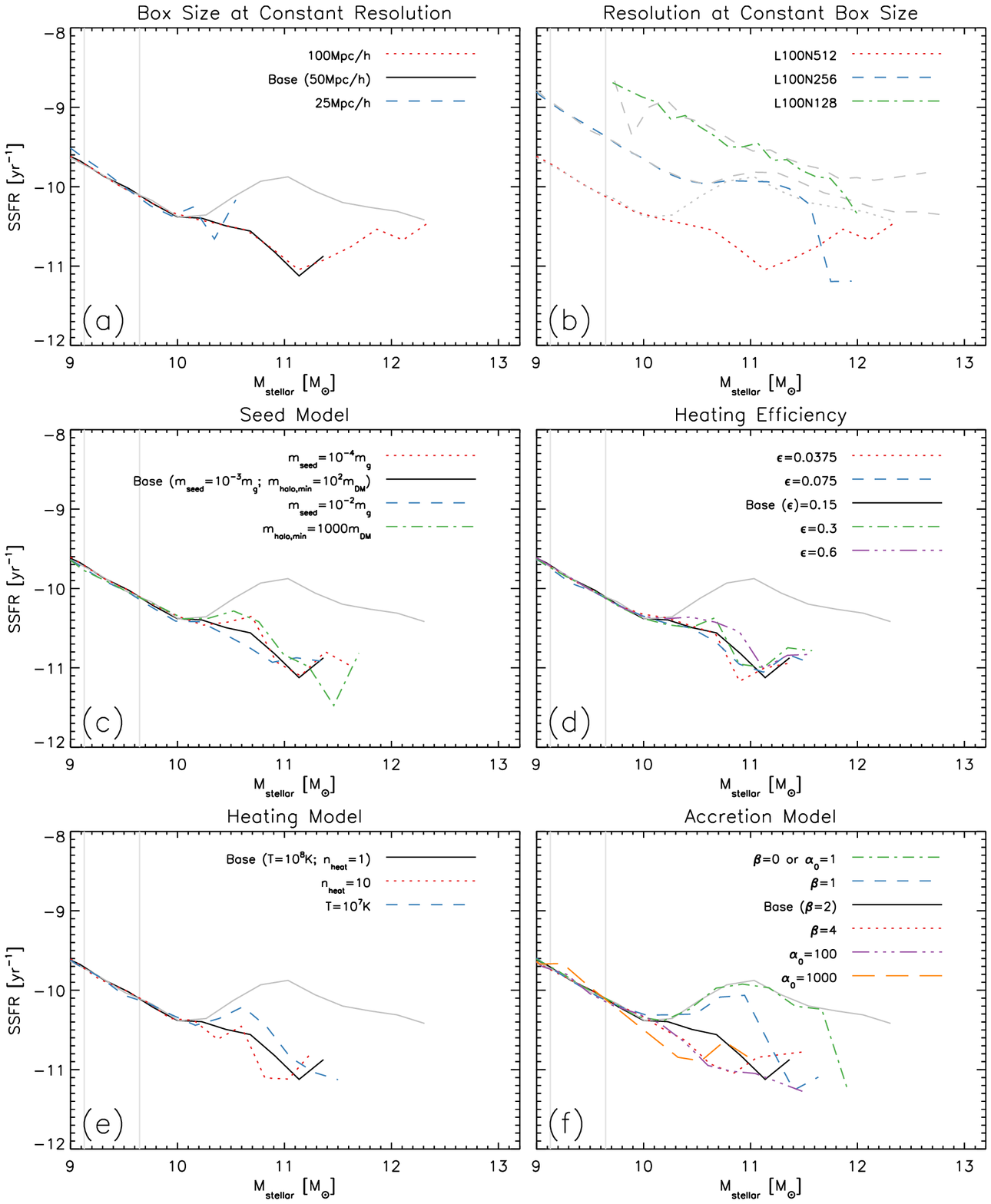}
\end{center}
\caption{The median SSFR as a function of galaxy stellar mass for each
  set of simulations. Line styles and panels correspond to those in
  Fig.~\ref{fig:params_sfr}.  The pale, grey curve in each plot
  represents the SSFR as measured in the simulation without AGN
  feedback (\emph{L050N256NOAGN}).  The solid black curve in each panel
  represents the fiducial simulation (\emph{L050N256}).  The upturn in
  galaxy SSFR in the simulations without AGN feedback at $M_*\approx
  3\times10^{10}\,{\rm M}_\odot$ is due to SN feedback becoming
  inefficient.  The two vertical lines in each panel indicate, for
  the fiducial simulation, the stellar mass above which 50\% and
  90\% of haloes contain a BH massive enough to have performed at
  least one heating event.}
\label{fig:params_ssfr}
\end{figure*}

We now turn our attention to the effect of varying each of the
parameters of our AGN model away from those selected in
Sec.~\ref{sec:pars}.  In order to quantify the effect of different
aspects of the AGN model on the results of our simulation, we split
the model parameters into four separate categories: the accretion
model ($\alpha_0$; $\beta$); the seed generation model ($m_{{\rm halo,
    min}}$; $m_{{\rm seed}}$); the feedback efficiency ($\epsilonf$);
and the heat distribution model ($\Delta T_{{\rm min}}$; $n_{{\rm
    heat}}$).  We look separately at the effects of changes in each of
these parameter sets, and additionally consider two purely numerical
effects: the simulation mass resolution and box size.  For each set of
simulations we make four diagnostic plots: in
Fig.~\ref{fig:params_sfr} we show the cosmic SFR density as a function
of redshift; Fig.~\ref{fig:params_bhdens} shows the evolution of the
global BH density, and the cumulative BH density present in seed-mass
BHs (grey curves); Fig.~\ref{fig:params_mbhmhalo} shows the redshift
zero $\mbh-M_{{\rm *}}$ and $m_{{\rm BH}}-\sigma$ relations.  We
associate BHs with gravitationally bound objects by identifying bound
substructures in the simulation using the algorithm {\sc subfind}
\citep{spri01b,dola08}.  We note that in this plot we show
\emph{total} halo stellar mass as a function of BH mass, as opposed to
the observations, where only the bulge stellar mass is calculated.
This means that all curves can be shifted slightly to the left.
Finally, Fig.~\ref{fig:params_ssfr} shows the median specific SFR
(SSFR) in bins of stellar mass.  In this plot, the grey lines
represent results from simulations that do not include AGN feedback.
In figures \ref{fig:params_ssfr} and \ref{fig:params_mbhmhalo} the
vertical lines represent the halo stellar masses at with 50\% and 90\%
of haloes contain BHs massive enough to have performed at least one
heating event.  It is immediately clear from
Fig.~\ref{fig:params_mbhmhalo} that the $\mbh-\sigma$ relation is much
more robust to changes in parameters than the $\mbh-M_*$ relation.

Each set of simulations is compared to our fiducial simulation
(\emph{L050N256}), which uses the model parameters that were justified in
Sec.~\ref{sec:pars}.  To aid comparison between the different
simulation sets, the fiducial simulation appears in every plot as a
solid, black curve.  Details of all the simulations discussed in this
section appear in Table \ref{tab:sims}.  We now discuss each
simulation set in turn.

\subsubsection{The effect of box size and mass resolution}

We consider first the effect of changing the box size at a constant
resolution by comparing models \emph{L100N512}, \emph{L050N256} and
\emph{L025N128}.  The size of the simulation box has a negligible
effect on both the star formation rate density
(Fig.~\ref{fig:params_sfr}a) and, for $z<4$, on the global mass of BHs
(Fig.~\ref{fig:params_bhdens}a).  Because the properties of individual
BHs are set by local physical processes, increasing the box size does
nothing to the scaling relations (Fig.~\ref{fig:params_mbhmhalo}a),
except for allowing us to probe the mass function up to larger halo
masses.  The same holds true for the SSFRs of individual objects
(Fig.~\ref{fig:params_ssfr}a).

We now assess the impact of numerical resolution on our results.  We
compare simulations at three different resolutions (\emph{L100N512},
\emph{L100N256}, and \emph{L100N128}) both with and without AGN
feedback.  Simulation $L100N128$ has a dark matter particle mass of
$3.54\times10^{10}\,{\rm M}_{\odot}$, a factor of 64 worse than that
used in \emph{L050N256}, our lowest resolution production simulation.

We concentrate first on the star formation history in these
simulations (Fig.~\ref{fig:params_sfr}b). At high redshift ($z>4$),
the star formation rate is governed by numerical resolution.  At low
redshift ($z<2$) the two highest resolution simulations
(\emph{L100N512} and \emph{L100N256}) have star formation rates that
differ by $\sim0.2\,$dex, indicating that this quantity is not yet
fully resolved.  We see by comparing simulations with and without AGN
that the factor by which AGN feedback decreases the global SFR is the
same in both simulations.  However, in the lowest resolution
simulation (\emph{L100N128}; blue, dashed line), the AGN are largely
ineffective at decreasing the global SFR.

We now turn our attention to the BH properties.  The $z<1$ integrated
BH density (Fig.~\ref{fig:params_bhdens}b) is virtually
indistinguishable between runs \emph{L100N512} and \emph{L100N256}.
This occurs because the global black hole density is dominated by the
massive BHs, which are well resolved in both simulations.
\emph{L100N128} underpredicts the redshift zero BH density by a factor
of two.  This is due to two reasons. Firstly, in this simulation seed
mass BHs are placed only in objects with masses larger than
$3.54\times10^{12}\,{\rm M}_\odot$, which means that we neglect a
large population of important black holes.  Secondly, we see that the
first BHs in this simulation start growing only at low redshift
($z\approx4$), and as such, may not have had enough time to grow onto
the observed BH scaling relations.

This picture is borne out by the properties of individual
BHs. Fig.~\ref{fig:params_mbhmhalo}b shows that for $\sigma>100\,{\rm
  km/s}$ the BH properties are well converged, whereas at the lower
mass end of the scaling relations the resolution
becomes important.  The lowest resolution simulation underpredicts the
BH scaling relations at all masses.

If we now examine the SSFRs of individual objects
(Fig.~\ref{fig:params_ssfr}b), looking first at simulations that do
not contain AGN (\emph{L100N512NOAGN} and \emph{L100N256NOAGN}; grey,
dotted and solid lines respectively) we see that the SSFR is almost
converged above a stellar mass of $10^{10.5}\,{\rm M}_\odot$.  If we
now consider the effect of adding AGN feedback to these simulations,
we see that the stellar mass at which the simulations with and without
AGN diverge from one another has not yet converged, indicating that
results on the scale of individual objects in these simulations may
remain affected by numerical resolution.  We note that the redshift
zero global SFR is affected by resolution to a lesser degree than the
SSFR.  This is likely to be because the total galaxy stellar mass is
an integral over all time of the galaxy star formation rates, so the
strong resolution effects at high redshift
(Fig.~\ref{fig:params_sfr}b) persist in the redshift zero population.

We thus conclude that, because local processes govern the size of
individual BHs, the simulation box size is unimportant when discussing
BH scaling relations.  However, the limited simulation mass resolution
means that the stellar masses in our simulated galaxies are not yet
fully resolved.  We also conclude that BH properties, such as the
integrated cosmic mass density and BH scaling relations are converged
in all of our production simulations, but decreasing the mass
resolution by a factor of 64 places us in the regime where no AGN
properties are resolved.  Therefore, we can conclude that the global
results we present in this paper are robust to changes in numerical
resolution, but that predictions involving the stellar properties of
individual objects should be treated with caution.

\subsubsection{The seed model}
\label{sec:seedmodel}
Two parameters control the way that seed BHs are generated: the BH
seed mass ($m_{{\rm seed}}$) and the halo mass above which BH seeds
are inserted into haloes ($m_{{\rm halo,min}}$).  For the minimum halo
mass we chose to use a fixed number of particles, $m_{{\rm
    halo,min}}=100m_{{\rm DM}}$, rather than a fixed halo mass because
this ensures that BHs are always placed into well defined haloes,
regardless of the numerical resolution.  Consideration of the AGN
feedback process then allowed us to place a minimum value on the
possible mass of BH seeds ($10^{-3}\mg$) such that the average time
between heating events for a BH accreting at the Eddington rate is
shorter than the Salpeter time, making it possible for the BH to
regulate its growth. We also noted that the requirement that the total
mass in seed mass BHs generated in a simulation is much lower than the
observed redshift zero cosmic BH density, does not allow the seed mass
to be much larger than the minimum possible value.  We thus chose
$m_{{\rm seed}}=10^{-3}\,m_{{\rm g}}$ as our fiducial value.  We now
examine how changes in these two parameters affect our results.
 
By comparing our base model (\emph{L050N256}) to a simulation in which
BH seeds are only placed into haloes that are ten times more massive
(\emph{L050N256HIHALO}; green dot-dashed curves) we can say that most
of our results are insensitive to the choice of $m_{{\rm halo,min}}$.
The global star formation rate density in the simulations
(Fig.~\ref{fig:params_sfr}c) is virtually unaffected by changing
$m_{{\rm halo,min}}$ from $10^2\,m_{{\rm DM}}$ to $10^3\,m_{{\rm
    DM}}$, probably because the SFR of galaxies is only affected by
AGN in the most massive haloes ($M_*\ga 10^{11}{\rm M}_\odot$;
Fig.~\ref{fig:params_ssfr}).  The cosmic BH density is also
insensitive to $m_{{\rm halo,min}}$.  This follows because in the
fiducial model, the cumulative seed BH mass makes up only $\sim2\%$ of
the redshift zero BH density, so removing some seeds has a negligible
effect on the global density.  We now turn our attention to the BH
scaling relations (Fig.~\ref{fig:params_mbhmhalo}c), where we see that
the BHs grow on to the observed scaling relations at a higher mass,
but that for $M_*\ga 10^{11}\,{\rm M}_\odot$ or $\sigma>100$~km/s the
results are almost identical to the fiducial simulation.  These are
the BHs that are energetically important and we can therefore conclude
that results derived from our AGN simulations are insensitive to
uncertainties in the minimum halo mass that contains a BH.

Large changes in the mass of the seed BHs (\emph{L050N256HISEED},
\emph{L050N256LOSEED}) can lead to more significant changes in the
global properties of the simulation. If the initial seed mass is
lowered by a factor of ten then by redshift zero the global BH density
is slightly greater than in the fiducial case
(Fig.~\ref{fig:params_bhdens}c).  This occurs because initially
smaller BHs take longer to grow onto the BH scaling relations both
because they need to grow more and because they grow more slowly (see
Fig.~\ref{fig:bhga}).  As a result of this the galaxy potential well
is deeper by the time that the BH begins to heat gas, and so it
requires more energy input for the BH to stop its own growth.  Because
it takes longer for AGN feedback to become effective, the global SFR
is higher (Fig.~\ref{fig:params_sfr}c).  The same argument can be made
in reverse for high seed masses and explains why increasing the BH
seed mass slightly decreases the global star formation rate
(Fig.~\ref{fig:params_sfr}c). We can draw the same conclusions from
examining the SSFR of individual objects
(Fig.~\ref{fig:params_ssfr}c). Changing the seed mass has virtually no
effect on the SFR of the most massive objects, and only affects haloes
near the threshold mass at which AGN feedback begins to become
important.

The cumulative amount of seed BHs in the simulation
\emph{L050N256HISEED} is within a factor of four of the redshift zero
black hole density.  Although this does not violate our constraint
that the total mass in seed BHs should be less than the observed
redshift zero value, it is barely satisfied, and leaves very little
room for AGN in our simulations to grow.  Additionally a large value
for the initial seed mass means that we place seed BHs initially
\emph{above} the BH scaling relations
(Fig.~\ref{fig:params_mbhmhalo}c).  Nevertheless they still grow onto
the same scaling relations.

We note that all our global results are a very weak function of the
seed mass. For example, changing the seed mass by a factor of 100
changes the global SFR by no more than a factor 2.5.  Other
uncertainties, particularly those introduced by our lack of knowledge
about the way in which BHs accrete, lead to us conclude that the
specifics of the seed model are relatively unimportant.

\subsubsection{The feedback efficiency}

The efficiency with which a BH's radiated energy is coupled to the
ambient gas, $\epsilonf$, is treated as a free parameter in our
simulations.  We show in this section that this parameter controls the
normalization of the total mass in BHs and of the BH scaling relations
\citep[see also][]{dima05}. In our simulations $\epsilonf$ was tuned 
to match the $z=0$ $\mbh-m_{{\rm halo}}$ relation and the redshift
zero cosmic black hole density.  A value of $\epsilonf=0.15$ was found
to work well.

Changing $\epsilonf$ from its fiducial value has no discernible effect
on the global star formation rate (Fig.~\ref{fig:params_sfr}d), or on
the SSFRs of individual objects (Fig.~\ref{fig:params_ssfr}d), but the
total redshift zero BH density is inversely proportional to the
feedback efficiency (Fig.~\ref{fig:params_bhdens}d).  Most strikingly,
simulations \emph{L050N256VHIEPS}, \emph{L050N256HIEPS},
\emph{L050N256LOEPS}, and \emph{L050N256VLOEPS} have values of
$\epsilon_{{\rm f}}$ that differ by factors of 4 , 2, 1/2, and 1/4
from the fiducial run, and the ratio of their $z=0$ BH densities to
that of the fiducial simulation are 0.24, 0.53, 1.93 and 4.03
respectively.  The fact that the final mass of BHs is directly
proportional to $1/\epsilonf$ implies that in our models any given BH
grows until it has injected a specific amount of energy per unit halo
mass, at which point it is able to reduce its local density and to
effectively self-regulate its growth.  As demonstrated in
\citet{boot09} this remarkable result agrees qualitatively with the
ideas presented in e.g.~\citet{fabi99} and \citet{silk98}, where BHs
grow until they can expel gas from the galaxy, at which point they
enter a quiescent phase.

The redshift zero global BH density (Fig.~\ref{fig:params_bhdens}d)
and the normalization of the BH scaling relations
(Fig.~\ref{fig:params_mbhmhalo}d) are both directly proportional to
$1/\epsilonf$, and so we can use these observations to constrain the
value of $\epsilonf$ that allows our model to be consistent with
redshift zero observations. We employ a value of $\epsilonf=0.15$ in
the present study, but note that we will show elsewhere that the
parameters of the AGN model must be tuned in conjunction with the SN
feedback prescription.

\subsubsection{The heating mechanism}

The resolution of cosmological simulations is too poor to resolve the
scales on which AGN inject energy into the ISM and as such we are
forced to make some assumptions.  It is therefore important to verify
that the results obtained from our simulations do not depend strongly
on the implementation of the energy coupling mechanism.

Two parameters control how energy is distributed from the AGN to its
surroundings: the minimum temperature increase and the number of
neighbouring SPH particles to heat ($\Delta T_{{\rm min}}=10^{8}$~K
and $\nheat=1$). The fiducial value of the minimum temperature
increase was constrained by two competing effects.  Firstly, if
$\Delta T_{{\rm min}}$ is too low, the cooling time of heated gas will
be so short that the gas will be able to radiate away its energy
before having a dynamical effect.  On the other hand, if $\Delta
T_{{\rm min}}$ is set too high, then the amount of energy needed to
perform a heating event will be so high that the numerically
determined time between heating events exceeds the Salpeter time-scale,
making it impossible for the BH to regulate its growth.  We found that
$\Delta T_{{\rm min}}=10^8$~K provides a good compromise between these
two effects.  Similarly, to minimize the numerically controlled duty
cycle we set $n_{{\rm heat}}=1$ in our fiducial models.  Here we
assess the impact of distributing feedback energy in different ways.

If we change either the temperature to which gas is heated from
$10^8$~K to $10^7$~K (\emph{L050N256T7}) or the number of black hole
neighbours affected by each feedback event from 1 to 10
(\emph{L050N256HINHEAT}), we see small changes in the cosmic star
formation history (Fig.~\ref{fig:params_sfr}e) and in the cosmic BH
density (Fig.~\ref{fig:params_bhdens}e).  We note, however, that the
results are relatively insensitive to these parameters, and that
changing the value of either $\Delta T_{{\rm min}}$ or $\nheat$ by an
order of magnitude affects both BH properties and SFRs by only
$\sim0.3$~dex.  We thus conclude that as long as the feedback
efficiency ($\epsilonf$) is calibrated such that the redshift zero
black hole relations are satisfied, and that the minimum heating
temperature is sufficient for feedback to be effective, our other
results are robust to the precise way in which this energy is coupled
to the ISM.

\subsubsection{The accretion model}

Because numerical simulations cannot resolve the properties of the
multi-phase ISM, we can justify the use of high values for the
multiplicative factor, $\alpha$, in the Bondi-Hoyle accretion rate
(see Eq.~\ref{eq:bhl}), at least for high density gas
(Sec.~\ref{sec:pars}).  This provides a motivation for the
constant-$\alpha$ accretion model -- similar to those previously
published -- where BHs accrete with a very high efficiency.  In this
paper we have introduced a new class of models that treat accretion of
low-density gas differently.  By noting that the ISM will not contain
a cold ($T\ll 10^4$~K) phase at low pressures and that we are also
able to resolve Bondi-Hoyle accretion onto BHs with masses greater
than the resolution limit of our simulation if we resolve the Jeans
mass, we argue that $\alpha$ must asymptote to unity in low density
environments.  We then parametrize our ignorance about the state of
the multiphase ISM, and the precise mechanism by which AGN accrete by
introducing a parameter, $\beta$, that describes a power-law
dependence of the accretion rate on the local gas density
(Eq.~\ref{eq:beta}).

The growth of a black hole depends upon the accretion model used.  In
the constant-$\alpha$ model (Bondi-Hoyle with $\alpha_0=100$) the
growth is Eddington limited unless the gas in the immediate
surroundings of the BH has a density that is much lower than typical
of the ISM, e.g.\ as a result of feedback from the BH.  In the
constant-$\beta$ models, however, much larger densities are required
for the accretion rate to become Eddington limited.  Efficient growth
is only possible if the BH's local density is enhanced by dynamical
processes.  BHs can only decrease their accretion rate (and hence
regulate their own growth) when densities are low enough that
accretion rates are no longer Eddington limited.  The density at which
BH accretion rates become Eddington limited depends on the accretion
model (Fig.~\ref{fig:rhoedd}).  In the constant-$\alpha$ model with
$\alpha_0=100$ BHs self-regulate at densities below the density at
which a multi-phase medium is expected to form
(Fig.~\ref{fig:rhoedd}), this has a large effect on the physical
properties of the galaxy and because the simulations resolve
Bondi-Hoyle accretion in this regime it invalidates the assumptions
used to justify the use of $\alpha_0=100$ in the first place.

We now consider how changes in the accretion model affect the
properties of the simulated galaxy population.  Increasing $\beta$
from 2 to 4 (simulation \emph{L050N256B4}) depresses the global star
formation rate somewhat (Fig.~\ref{fig:params_sfr}f) because the
$\beta$ parameter controls the gas density at which the accretion rate
for a given BH mass becomes Eddington limited (Fig.~\ref{fig:rhoedd}).
Hence, for a given value of $\beta$ there is a critical value of the
local gas density above which a BH can \lq switch on\rq\, and begin to
grow rapidly.  For larger values of $\beta$ this occurs at a lower
density, and hence in smaller haloes.  This manifests itself in the BH
scaling relations by changing the minimum galaxy mass at which the BHs
grow onto the observed scaling relations.  A lower value of $\beta$
therefore increases the global star formation rate by allowing BHs to
grow only in more massive haloes.  Increasing $\beta$ above 4 does not
have a large effect on any of our results, as for any large value of
$\beta$, the BH accretion model behaves in such a way as to be
Eddington limited in star-forming gas (Fig.~\ref{fig:rhoedd}), and the
difference in behaviour between a $\beta=4$ model and a $\beta=\infty$
model is very small.  Any physical process that is strongly dependent
on the local density is affected by numerical resolution, because
higher resolution simulations allow the formation of higher density
regions.  For this reason we caution that the stellar mass at which
BHs grow onto the observed relation is affected by numerical
resolution, and so the value of $\beta$ may need to be tuned to
different values for simulations with mass resolutions significantly
different to those presented here.

A comparison of the effect of different $\beta$ models on the SSFR of
haloes (Fig.~\ref{fig:params_ssfr}f) supports this picture.  The
stellar mass above which the behaviour of each AGN model diverges from
the behaviour of the simulation without AGN depends upon the value of
$\beta$.

In the constant-$\alpha$ model with $\alpha_0=100$
(\emph{L050N256A100B0}) BHs grow in an Eddington limited manner from
their birth, and as such suppress star formation in all haloes that
contain a BH.  This is visible in Fig.~\ref{fig:params_ssfr}f, in
which the constant-$\alpha$ simulation deviates from the simulation
without AGN in haloes with a low stellar mass.  The constant-$\alpha$
simulations efficiently suppress star formation in every halo that
contains a BH and as such the integrated SFR is more than an order of
magnitude lower than in the other simulations.  The same effect is
present in the large simulation volume, but is less pronounced because
seed mass BHs are placed only into larger haloes, where both accretion
models are capable of suppressing SF.  This result implies that in
order for a simulation that employs a constant-$\alpha$ accretion
model to reproduce the observed break in galaxy properties at $\log
(M_*)\sim10.5$ \citep[e.g.][]{kauf03}, the resolution must be tuned
such that BH seeds are placed into haloes of the correct
size\footnote{Because the accretion rate also depends on the assumed
  effective EOS, this may not be true for all constant-$\alpha$ models
  used in the literature. In the models of S05, which employ an EOS
  that is initially stiffer than our $\gamma_{\rm eff}=4/3$, the BH
  growth time has a local minimum at $n_{{\rm H}}=n_{{\rm H}}^*$ which
  suppresses AGN growth relative to our constant-$\alpha$ model
  (Springel, private communication).}.

Considering now reducing the value of $\alpha_0$ in the
constant-$\alpha$ models from 100 to 1 (\emph{L050N256A100B0} compared
to \emph{L050N256B0}; a constant-$\beta$ model with $\beta=0$ is
equivalent to a constant-$\alpha$ model with $\alpha_0=1$), equivalent
to removing the numerical \lq fudge factor\rq\, from the BH accretion
rates, we see that BHs are unable to grow in all but the very most
massive haloes (Fig.~\ref{fig:params_mbhmhalo}f), and so the halo
SSFRs are virtually the same as for the simulation without AGN
feedback in all haloes up to a stellar mass of $10^{12}\,{\rm
  M}_\odot$ (Fig.~\ref{fig:params_ssfr}f).  This, in turn, means that
the global SFR density in the $\alpha_0=1$ simulation is very similar
to that in the simulation run without AGN
(Fig.~\ref{fig:params_sfr}f).  The global BH density at redshift zero
is actually higher than that in the fiducial simulation
(Fig.~\ref{fig:params_bhdens}f).  This occurs because the BHs that do
grow in the $\alpha_0=1$ simulation grow very late, and so find
themselves in a deeper potential well.  In order to self-regulate they
then need to output more energy and hence grow even more than in the
other simulations.  The global BH density is therefore by itself not a
good indicator of how efficiently AGN are able to suppress star
formation.

We now consider increasing $\alpha_0$ by an order of magnitude
relative to the usual value (\emph{L050N256A1000B0} compared to
\emph{L050N256A100B0}).  We already know that the density below which
BHs can self regulate depends strongly on the value of $\alpha_0$
(Fig.~\ref{fig:rhoedd}), and it follows that the value of $\alpha_0$
will have a strong effect on the physical conditions in the centres of
haloes.  In the case of $\alpha_0=1000$ the global BH density is
similar to the $\alpha_0=100$ case (Fig.~\ref{fig:params_bhdens}f),
whereas the global SFR and galaxy SSFRs are much lower than in any
other simulation (Fig.~\ref{fig:params_sfr}f and
Fig.~\ref{fig:params_ssfr}f), and the $m_{{\rm BH}}-m_{{\rm stellar}}$
relation is shifted significantly to the left as the BHs effectively
shut down star formation in all haloes.  However, the $m_{{\rm
    BH}}-\sigma$ relation is not as strongly affected by the very
effective AGN feedback.  This is likely due to the fact that the
stellar velocity dispersion tracks the depth of the galaxy potential
well, which contains a large contribution from dark matter, whereas
the galaxy stellar mass is sensitive to the details of the feedback
processes operating inside the galaxy.  We see that changing the value
of the free parameter $\alpha_0$ can have profound effects on the
simulated galaxy population, even if the evolution in the global mass
density of BHs is barely affected by the parameter change.

We conclude this section by noting that predictions from AGN feedback
models of this type are sensitive to the accretion model that is used.
More generally, it is not clear that the Bondi-Hoyle rate is the
correct accretion rate to use in the case of a BH accreting from a hot
halo \citep{krum06}.  We find that a different parametrization of the
accretion rate leads to profound differences in the star formation
histories and speed of BH growth and therefore caution the reader that
the AGN accretion mechanism represents a significant source of
uncertainty in all our results.

%
%
\section{Discussion and conclusions}
\label{sec:discussion}

We have presented and tested a method to incorporate SMBHs into
cosmological, smoothed particle hydrodynamics simulations. The method,
which is a substantially modified version of the one introduced by
S05, self-consistently describes the mass growth of BHs and feedback
from AGN. Here we consider growth through mergers with other BHs as
well as through accretion of gas. The AGN feedback in our model is
thermal and local to the BH.

Although we also use the SPH code {\sc gadget III}, our code differs
from that of S05 in many ways, including the use of different models
for star formation, feedback from SN, radiative cooling and stellar
evolution. Particularly relevant for AGN feedback is the fact that,
contrary to S05, we do not make use of a subgrid model to describe the
different phases of the ISM.  Following S05, we make use of subgrid
BHs to allow BH masses to be small compared with the masses of the
particles containing the BHs. We note, however, that while this
approach allows one to significantly extend the range of BH masses in
the simulation, the accretion radius will only be resolved if the BH
mass exceeds the local Jeans mass. Unfortunately, this is generally
not the case if the BH mass is small compared with the particle mass.

The AGN model is specified by seven main parameters (Table
\ref{tab:parlist}). Two parameters describe the BH seed generation
mechanism: the BH seed mass, $m_{\rm seed}$, and the halo mass into
which seed BHs are placed, $m_{\rm halo,min}$. Two parameters describe
the amount of energy that is coupled back to the ISM per unit accreted
mass: $\epsilon_{\rm r}$, the radiative efficiency of a BH accretion
disk, is the fraction of the rest mass energy accreted by the BH that
is radiated by the AGN and $\epsilon_{\rm f}$ is the fraction of the
radiated energy that is coupled thermally to the ISM. For a given BH
accretion rate, the rate at which energy is injected into the ISM
depends only on the product $\epsilon_{\rm r}\epsilon_{\rm
  f}$. However, we do need to specify the radiative efficiency
$\epsilon_{\rm r}$ since the mass growth of the BH is proportional to
$(1-\epsilon_{\rm r})$.  A further two parameters control the
numerical implementation of the injection of energy into the ISM by
AGN: the number of neighbouring gas particles heated by each BH
heating event, $\nheat$, and the minimum amount by which their
temperature is increased, $\Delta T_{{\rm min}}$. Because we let BHs
store feedback energy until they have saved enough to heat $\nheat$
particles by $\Delta T_{{\rm min}}$ degrees Kelvin, these last two
parameters together determine the AGN duty cycle for a fixed accretion
rate. Finally, we require one additional parameter that controls how
BHs accrete gas, and we describe two different models for this
process.

The gas accretion rate is assumed to scale as the
Bondi-Hoyle accretion rate, evaluated at the location of the BH and on
the scales resolved by the simulation. We do not, however, allow the
accretion rate to exceed the Eddington rate. The Bondi-Hoyle
accretion rate predicted by the simulation will 
underestimate the true rate if the density is
underestimated or if the temperature is overestimated. A lack of
numerical resolution may result in an underestimate of the gas
density, which motivated S05 to multiply the
Bondi-Hoyle rate predicted by the simulation by a constant factor
$\alpha_0=100$. We 
call models of this type constant-$\alpha$ models. We noted that
cosmological simulations lack not only the resolution, but also the
physics to model the cold ISM phase. For example, our simulations use
a polytropic effective equation of state for densities at which 
the gas is expected to be multiphase. Hence, they will miss the cold
component for which the Bondi-Hoyle accretion rate would be
highest. This will lead us to strongly underestimate 
the Bondi-Hoyle rate in high density gas.

Although cosmological simulations cannot yet resolve cold,
interstellar gas, many do resolve the Jeans scales at densities low
enough for the ambient ultraviolet radiation to suppress
cooling below $10^4\,$K. Specifically, any simulation that resolves
the Jeans scales in the gas surrounding a BH particle, will also
resolve the Bondi-Hoyle radius if the BH mass exceeds the local Jeans
mass. Hence, at sufficiently low densities the Bondi-Hoyle accretion
rate is modeled correctly and multiplying it by $\alpha_0=100$ would
result in a large overestimate of the accretion rate. 

We therefore introduced a second class of models in which the
Bondi-Hoyle accretion rate is multiplied by a factor $(n_{\rm
  H}/n_{\rm H}^\ast)^\beta$ for densities $n_{\rm H} > n_{\rm
  H}^\ast$, where $n_{\rm H}^\ast$ is the density above which the gas
is expected to be multiphase (we take $n_{\rm H}^\ast=0.1~{\rm
  cm}^{-3}$). We refer to this class of models as constant-$\beta$
models. Note that both constant-$\alpha$ and constant-$\beta$ models
use a single free parameter. Because we have changed the
density-dependence of the accretion rate, we cannot claim to be
simulating Bondi-Hoyle accretion, even though the changes are
motivated by the Bondi-Hoyle formula and even though we do retain the
Bondi-Hoyle scaling with the BH mass. We argued, however, that this is
also true for constant-$\alpha$ models because the use of values
$\alpha_0 \gg 1$ implies that the densities and temperatures predicted
by the simulations are greatly in error.

The parameters $\alpha_0$ and $\beta$, used in the constant-$\alpha$
and constant-$\beta$ models respectively, control the ambient gas
density at which the BH accretion rate becomes Eddington limited
(Fig. \ref{fig:rhoedd}). Because the maximum densities sampled by the
simulation increase with halo mass (both because more massive haloes
are resolved with more particles and because the central pressure
increases with the depth of the potential), these parameters
effectively set the halo mass above which BHs can grow efficiently.
We set $\beta=2$, which we find results in efficient BH growth in
haloes with stellar masses $\ga 10^{10.5}\,{\rm M}_\odot$ in these
simulations. Using a constant-$\alpha$ prescription with
$\alpha_0=100$ implies that, in the absence of AGN feedback, the
accretion rate is essentially always Eddington limited. Because the
accretion is efficient even at relatively low gas densities, AGN
feedback is in that case important in all haloes that exceed $m_{\rm
  halo,min}$. For this class of models the halo mass above which AGN
can suppress star formation is thus in effect a free parameter
($m_{\rm halo,min}$).  For BHs with masses greater than $10^6\,{\rm
  M}_\odot$ self-regulation can only occur at densities orders of
magnitude below the star formation threshold (Fig.~\ref{fig:rhoedd}).
In this regime we resolve Bondi-Hoyle accretion, invalidating the
assumption used to justify large values of $\alpha$ in the first
place. Constant-$\alpha$ models therefore underestimate the gas
density required for self-regulation and will thus overestimate the
suppression of the star formation rate.

Having chosen a prescription for gas accretion, we then derive values
for the other model parameters. Because each BH stores its feedback
energy until it suffices to heat $n_{\rm heat}$ neighbours by $\Delta
T_{\rm min}$, we are faced with a numerically determined duty cycle
(for a given accretion rate). In order for the BH to be able to
regulate its growth, we require the time between heating events,
$\Delta t_{\rm heat} \propto n_{\rm heat} \Delta T_{\rm min}$, to be
as small as possible and certainly smaller than the Salpeter time if
the accretion is Eddington limited. We use $\Delta T_{{\rm
    min}}=10^{8}\,$K, which ensures the temperature of the heated gas
is high enough that the injected thermal energy is not just radiated
away, and $\nheat=1$, which minimizes $\Delta t_{\rm heat}$. Because
$\Delta t_{\rm heat}$ decreases with the ratio of the mass of the BH
to that of the heated gas particle, the requirement that $\Delta
t_{\rm heat}$ is smaller than the Salpeter time for Eddington-limited
accretion implies that the (subgrid) BH mass must exceed 0.1 per cent
of the gas particle mass (Fig.~\ref{fig:bhg}). Hence, we set $m_{\rm
  seed}=10^{-3}\,\mg$.  We set $m_{{\rm halo,min}}=100m_{{\rm DM}}$ in
order to ensure that seed BHs are placed only into well defined
haloes. These parameter values will obviously need to be increased if
they result in seed and/or minimum halo masses that are lower than
expected physically, as may be the case for simulations that use a
much higher resolution than is used here.

We assume the standard value $\epsilon_{\rm r}=0.1$ for the radiative
efficiency and tune $\epsilon_{\rm f}$ to match the redshift zero
$\mbh-m_{{\rm halo}}$ relation and cosmic 
BH density.  A value of $\epsilonf=0.15$ was found to
provide a good match to the observations. 

Having specified the AGN model, we then analysed the results from a
large suite of 
cosmological simulations chosen to investigate the sensitivity of the
predictions on the model parameters. For this purpose we compared the
predictions for the cosmic SF history, the cosmic BH density, the
redshift zero BH scaling relations (both the $M_{\rm BH}-M_\ast$ and the
$M_{\rm BH}-\sigma$ relations), and the redshift zero galaxy specific star
formation rates (SSFRs). 

We demonstrated that the fiducial model provides good agreement with
both the observed mass density in BHs (Fig.~\ref{fig:params_bhdens}) and
the BH scaling relations (Fig.~\ref{fig:params_mbhmhalo}), and that
the inclusion of AGN feedback in the simulations effectively
suppresses star formation in galaxies with stellar masses greater than
$>10^{10.5}\,M_{{\rm odot}}$ (Fig.~\ref{fig:params_ssfr}).  We will
discuss the comparison between the simulated global SFR density and
observations elsewhere, but for now we note that the SN feedback parameters
of our models were tuned such that a simulation without AGN feedback
has a peak SFR density that is in good agreement with observations.
As such, adding an extra source of feedback energy inevitably results
in an underestimate of the SFR density. In order to achieve a good
match with observations the properties of the SN model must be tuned
in conjunction with those of the AGN model.

However, the focus of our study was not to match observations, but to
explore the dependence of the results on the parameters of the
model. Our main conclusions are summarised below:

\begin{itemize}
\item Regardless of whether BH seeds are initially placed above or
  below the BH scaling relations, they grow onto the same relations.

\item Because the global BH density is dominated by massive BHs and
  because AGN feedback is only important in high-mass haloes,
  uncertainties in the seed model employed do not lead to significant
  changes to the global properties of our simulations. Changing the
  initial seed mass by two orders of magnitude changes the global star
  formation rate by only a factor of 2.5.  The assumed seed generation
  model can, however, affect galaxy properties at around the galaxy
  mass where AGN first begin to become energetically important. At
  higher masses galaxy properties are largely insensitive to the
  initial distribution of BH seeds.

\item As discussed more comprehensively in \citet{boot09}, the
  normalization of both the global BH density and the BH scaling
  relations is nearly exactly inversely proportional to the AGN
  feedback efficiency, $\epsilon_{\rm r}\epsilon_{\rm f}$. Most
  strikingly, changing the efficiency by a factor 16 does not give
  rise to any discernible changes in the global SF history. These
  results imply that the total amount of thermal energy injected by
  AGN is conserved when the feedback efficiency is changed. These
  results can be explained if BHs grow until they have generated
  enough energy to regulate the accretion rate and if the required
  amount of energy depends on the depth of the potential well on
  scales that are larger than the radius on which the BH dominates.

\item Changing the way in which the thermal energy from the AGN is
  distributed in the gas surrounding the BH has little effect on our
  results, as long as the gas is heated to a temperature that is high
  enough for its cooling time to become long. Hence, thermal feedback
  can be efficient in cosmological simulations that do not resolve the
  multiphase ISM.

\item Cosmological simulations currently underestimate the Bondi-Hoyle
  accretion rate in dense gas because they lack both the resolution
  and the physics to model the dense, cold phase of the ISM. It is
  therefore necessary to increase the predicted accretion rate by a
  fudge factor, either by explicitly multiplying the accretion rate by
  a numerical correction factor or by employing a subgrid model for
  the unresolved gas physics to artificially boost accretion rates in
  star-forming gas.  Using a multiplicative factor that asymptotes to
  unity in the regime where the simulation is able to model the
  relevant physics (our constant-$\beta$ model) gives different
  results from using a high factor throughout (the constant-$\alpha$
  model) as has been used in most previous work. In general, the
  density above which BHs are able to accrete efficiently depends upon
  the accretion model used (Fig.~\ref{fig:bhga}). Because higher mass
  haloes contain higher density gas, the accretion model determines
  the halo mass above which AGN feedback becomes effective. Until the
  simulations are able to resolve Bondi-Hoyle accretion in a
  multiphase ISM, the predictions of the models will remain subject to
  significant uncertainty.

\item The $\mbh-\sigma$ relation is more robust than the $\mbh-M_*$
  relation to changes in the model parameters
  (Fig.~\ref{fig:params_mbhmhalo}), and so provides a test of the
  numerical model that is less affected by uncertainties in
  numerical parameters than the Magorrian relation.  This is likely
  due to the fact that the stellar velocity dispersion tracks the
  depth of the galaxy potential well, which contains a large
  contribution from dark matter, whereas the galaxy stellar mass is
  sensitive to the details of the feedback processes operating inside
  of the galaxy.

\end{itemize}

In summary, we have presented and tested a new model for the self-consistent
growth of BHs and feedback from AGN in cosmological simulations.  In a
future work we will discuss the interaction between AGN
feedback and other physical processes, and show that the results
obtained from an AGN model depend also on other processes
such as SN feedback and stellar mass loss.

\section*{Acknowledgments}
We are very grateful to Volker Springel for generously providing us
with a copy of his code. We would also like to thank him as well as
Richard Bower, Ian McCarthy and the members of the OWLS collaboration
for useful discussions. The 
simulations presented here were run on the Cosmology Machine at the
Institute for Computational Cosmology in Durham as part of the Virgo
Consortium research programme, on Stella, the LOFAR BlueGene/L system
in Groningen, and on Huygens, the Dutch national supercomputer. This
work was supported by Marie Curie Excellence Grant MEXT-CT-2004-014112
and by an NWO Vidi grant.

\label{lastpage}

\end{document}